# The Landscape of Generative AI in Information Systems: A Synthesis of Secondary Reviews and Research Agendas

Aleksander Jarzębowicz*, Adam Przybyłek, Jacinto Estima, Yen Ying Ng, Jakub Swacha, Beata Zielosko, Lech Madeyski, Noel Carroll, Kai-Kristian Kemell, Bartosz Marcinkowski, Alberto Rodrigues da Silva, Viktoria Stray, Netta Iivari, Anh Nguyen-Duc, Jorge Melegati, Boris Delibašić, Emilio Insfran
*Corresponding author: aleksander.jarzebowicz@pg.edu.pl

**Abstract**

**Context:** The post-ChatGPT surge has rapidly reframed Information Systems (IS) research and practice. As organizations and society grapple with Generative AI (GenAI) adoption, a body of secondary studies and research agendas has emerged to synthesize early evidence and chart directions for future inquiry.
**Objective:** This study conducts a systematic literature review of secondary studies and research agenda/roadmap papers to synthesize the state of knowledge on GenAI's benefits and challenges in IS, and to identify future research directions.
**Method:** We performed a systematic search across Scopus, Web of Science, and the AIS eLibrary for publications from 2023 onwards. Following a rigorous, multi-stage screening process, we selected a final set of 28 papers (18 secondary studies and 10 research agendas) for analysis using bibliometric mapping and thematic analysis. We also conducted a quality assessment of all sources to gauge confidence in each source's contribution to the findings.
**Results:** GenAI offers transformative potential to drive productivity, accelerate innovation, personalize services, and democratize access to expertise. However, its adoption is constrained by interrelated challenges: technical unreliability (hallucinations, performance drift), societal-ethical risks (bias, malicious misuse, skill erosion), and a governance vacuum (privacy, accountability, intellectual property).
**Conclusions:** Interpreted through a socio-technical lens, our findings reveal a persistent misalignment between GenAI's fast-evolving technical subsystem and the slower-adapting social subsystem, positioning IS research as critical for achieving joint optimization. To bridge this gap, we propose a research agenda that reorients IS scholarship from analyzing impacts toward actively shaping the co-evolution of technical capabilities with organizational routines, societal values, and regulatory institutions – emphasizing hybrid human-AI ensembles, situated validation, design principles for probabilistic systems, and adaptive governance. For practitioners and policymakers, responsible adoption requires balancing automation with human augmentation alongside transparent governance and adaptive regulations to ensure broadly shared benefits.

## 1. Introduction

The public release of ChatGPT in late 2022 was a decisive turning point in the evolution of artificial intelligence, triggering an unprecedented surge in the visibility and adoption of Generative Artificial Intelligence (GenAI) systems. In the following months, large language models (LLMs) and multimodal GenAI systems evolved from experimental technologies to widely deployed digital infrastructures, reshaping how information has been created, accessed, interpreted and acted upon across organizations and society [1–5]. Unlike earlier waves of AI that were largely constrained by narrow and task-specific applications [6], GenAI systems exhibit general-purpose capabilities that directly intersect with the core concerns of the information systems (IS) discipline, such as work practices, organizational processes, decision-making, governance, and socio-technical change [4, 5, 7–15].

For IS scholars, GenAI is not simply a continuation of prior AI research. It introduces a qualitatively different class of digital artifacts, namely probabilistic, generative, and conversational systems, that shift how system behavior is produced and evaluated and, in turn, blur boundaries between users and systems, automation and user interaction, and the production and consumption of knowledge [9, 16]. These systems are increasingly integrated into activities traditionally regarded as human-centric, including sense-making, creative work [11], software development [17–26], and professional judgment. Early evidence from knowledge-intensive institutional settings illustrates the implications for governance and adoption. In higher education, for example, GenAI adoption has raised concerns about academic integrity, critical thinking skills, academic standards, and implications for institutional policy and governance [27]. In healthcare, the extent to which patients adopt GenAI health assistants remains unclear [28]. Survey data suggests that trust and perceived benefits predict intention to adopt, while privacy concerns and resistance to change are associated with higher perceived risk [29]. Furthermore, IS scholarship has begun to frame GenAI as augmentation embedded in socio-technical systems rather than as an autonomous replacement [4, 9, 11, 13, 18, 26, 30], which raises questions about system agency, control, accountability, and the division of labor between humans and machines [11, 12, 25, 30–32]. Such questions are addressed by recent legal and governance initiatives [23, 30, 33, 34].

The rapid dissemination of GenAI has been accompanied by an equally rapid expansion of scholarly work [9]. For instance, in a remarkably short time span, the IS community and related disciplines have produced a substantial body of secondary studies, such as literature reviews, scoping reviews, mapping studies, as well as forward-looking research agendas and roadmaps [7, 8, 35]. Collectively, these works have aimed to synthesize early empirical evidence, propose conceptual frameworks, and reflect on directions for future research. However, the diversity of this literature has created a fragmented and difficult-to-navigate knowledge landscape. Individual reviews often focus on specific application domains (e.g., healthcare, education, software engineering), particular risk dimensions (e.g., bias, privacy, reliability), or narrow methodological perspectives, making it challenging to grasp an integrated understanding of the state of GenAI research within or related to the IS discipline [31]. GenAI adoption has shown a growing tension between technical capabilities and social readiness [9, 12–15, 30]. While the literature consistently highlights substantial benefits, such as productivity and personalization gains [26, 36], scalability, agility [37], and innovation [11, 14], it also reports profound challenges, including hallucinations and technical unreliability, ethical and societal risks, and unsolved governance and regulatory issues [2, 17, 26, 28, 29, 32, 38]. These tensions are not merely implementation issues; they reflect deeper misalignment between rapidly evolving technical systems and more slowly adapting social, organizational, and institutional structures [13, 28, 29]. They place GenAI within the intellectual core of IS as a socio-technical discipline concerned with the joint optimization of this current wave of technologies and social systems.

Therefore, conducting a systematic review in this domain is necessary but poses several non-trivial challenges. First, the GenAI literature is young and exceptionally fast-evolving, with conceptual vocabularies, application boundaries, and methodological conventions still in flux. This instability increases the risk of conceptual fragmentation, speculative claims, and overlapping or redundant syntheses. Second, existing secondary studies and research agendas vary considerably in scope, rigour, and epistemological orientation, ranging from tightly focused domain reviews to broad, visionary position papers. Integrating insights across such heterogeneous sources requires careful methodological design and transparent synthesis procedures. A third challenge lies in the inherently socio-technical nature of GenAI adoption: many reviewed studies span technology-centric narratives and socially oriented critiques, often without explicitly integrating these perspectives. As a result, benefits and risks are frequently discussed in isolation rather than as interdependent aspects of complex socio-technical systems.

Addressing this gap requires a synthesis approach that considers both technical and social dimensions. Thus, the primary goal of this research is to provide a coherent, field-level synthesis of how GenAI is currently being framed, adopted, and evaluated within the Information Systems scholarship, culminating in an integrative conceptual framework – the Sociotechnical GenAI Outcomes Matrix (SGOM) – that connects the benefits and challenges identified in the literature to concrete design, governance, and evaluation responses across levels of analysis. Specifically, this study has the following objectives:

- Map the landscape of secondary studies and research agenda papers on GenAI in the IS field;
- Synthesize the benefits and challenges identified in these works; and
- Identify and discuss research gaps and future directions as an integrated agenda that can be theoretically grounded and socially relevant for the IS community.

This study addresses these objectives by conducting a systematic literature review (SLR) of secondary studies and research agenda papers on GenAI in the IS domain published since 2023. By explicitly focusing on integrative contributions and agenda-setting contributions, the review captures how the IS community is collectively interpreting early evidence, diagnosing risks, and envisioning future research in this scope. Through a combination of bibliometric analysis and thematic synthesis, the review maps the structure of this emerging knowledge base, synthesizes benefits and challenges, and presents the research gaps and directions proposed in the identified literature.

The contribution of this study is fourfold. First, it provides a comprehensive and methodologically consistent overview of the secondary and agenda-setting literature on GenAI in IS, offering clarity in a rapidly expanding fragmented research space. Second, it synthesizes the dominant narratives surrounding GenAI's transformative potential and its associated risks, revealing persistent forces that cross application domains and methodological traditions. Third, building on this synthesis, we introduce the Sociotechnical GenAI Outcomes Matrix (SGOM), a novel conceptual framework that maps benefits, failure modes, design controls, and evidence/KPIs across six interdependent levels of analysis (user, organizational, societal, ethical, engineering, and quality), providing the IS community with a diagnostic and design-oriented instrument for evaluating and steering GenAI implementations toward durable, legitimate outcomes. Fourth, by aggregating and structuring the proposed research directions, the study articulates a consolidated research agenda that highlights critical opportunities for IS scholars to advance theory, inform practice, and even shape policy. By positioning GenAI as a socio-technical phenomenon rather than a purely technical innovation, this review underscores the IS discipline's distinctive role in shaping the responsible evolution of generative technologies.

The remainder of this paper is structured as follows. Section 2 provides essential background on GenAI and situates our study within the Information Systems perspective. Section 3 discusses related work. Section 4 details the method, including the research questions that guide our analysis, search strategy, selection, and analysis procedures. Section 5 presents the descriptive and thematic results of our analysis. Section 6 elaborates the synthesized benefits, challenges, and future directions. In Section 7, we interpret these findings through a socio-technical lens and discuss their broader implications. Section 8 translates the findings into the future research agenda. In Section 9, threats to validity and limitations are discussed. Finally, Section 10 concludes the article by summarizing our contributions and outlining key implications for research and practice.

## 2. Background

### 2.1. Generative AI

Since its inception in 1943, the concept of Artificial Intelligence (AI) has encompassed different paradigms, with the most relevant being the symbolic and the connectionist approaches [39, 40]. The symbolic paradigm is based on concepts and their relationships, i.e., inferential rules, that are used to perform reasoning [39]. While being effective for well-constrained domains, symbolic systems struggled with ambiguity, unstructured data, and open-ended tasks [31]. In contrast, the connectionist paradigm is based on large networks of simple processors, i.e., artificial neural networks (ANN), in which knowledge is encoded in the numerical strength of the connections between these processors [40]. These connections are optimized through a training process based on a set of inputs and expected outputs. In other words, an ANN approximates a (non-linear) mathematical function for which some inputs and outputs are given (training dataset). Initially, these processors, i.e., neurons, were grouped into layers that were connected in a single direction, i.e., the outputs of a layer were the inputs of the following layer. A limitation of these networks (feed-forward networks) was their stateless nature which limited their application for sequence analysis, such as natural language processing. The introduction of Recurrent Neural Networks (RNNs), in which the outputs of some layers could be used as inputs to previous layers, demonstrated the capacity to internally store a state, or memory, thereby obtaining better results for tasks demanding such capability [41]. However, early RNNs suffered from limited memory retention, consequently exhibiting degraded performance with long sequences. This limitation drove the development of Long Short-Term Memory (LSTM) networks [42]. In the last decade, the hardware improvements and reduced costs allowed the creation of larger, or deeper, networks which could be trained using the large amounts of data available in the Internet. These deep neural networks have been extensively explored to support software development [43].

Since neural networks represent mathematical functions, the analysis of text-based inputs, such as programming code, requires a conversion into numerical data, a process known as embedding. One way to perform this embedding is to represent words as vectors in a vector space. An example of this approach is Word2Vec [44], which captures syntactic and semantic relationships between words through training on large text corpora. However, because these early approaches mapped words to static vectors, they failed to capture surrounding context and struggled with polysemous words. ELMo [45] tackled these issues by employing an LSTM architecture.

A limitation of RNNs, including LSTMs, is their inherently sequential nature, which, for example, inhibits parallelization [46]. To tackle this issue, [46] proposed the transformer model, which became a key innovation for neural networks. Transformers rely on self-attention mechanisms that evaluate how each token in a sequence relates to others, enabling efficient parallelization and the capture of long-range dependencies. LLMs represent the most widely deployed class of transformer-based foundation models. The transformer is based on an encoder, which converts the complete input into an output of embeddings, and a decoder that, based on the previous tokens of the output, predicts the next output token. The stack of different layers of encoders and decoders led to different models, divided into encoder-only, decoder-only, and encoder-decoder models [47]. Encoder-only models, such as BERT [48], were the earliest LLMs and had good performance for natural language understanding [47]. However, decoder-only or autoregressive models, such as the GPT-series, obtained better performance in few-shot or even zero-shot generative tasks [49], especially after the inclusion of further training based on human feedback, as initially done for the InstructGPT by OpenAI [50]. This training approach allowed the launch of ChatGPT, in November 2022, that inaugurated the popularity of GenAI technologies.

Foundational models, such as BERT, GPT-4, Gemini, or LLaMa, are trained on large datasets for generic tasks, such as predicting the next token or masked token modelling, but they can be adapted to different downstream tasks. Initially, this process consisted on further training the model with a specific dataset, i.e., fine-tuning. For example, based on a dataset of vulnerable software functions, [51] fine-tuned the pre-trained CodeBERT model for the task of vulnerability detection. However, the few-shot or even zero-shot performance of decoder-only models has several advantages over fine-tuning, such as no need of larger datasets, knowledge and dedicated hardware for training these models, obtaining similar results in a fraction of the time. Therefore, it became important to identify techniques to prompt these models, i.e., prompt engineering, in a way to improve the quality of the output [14, 24, 52]. A particular approach that significantly improved the performance, especially for tasks requiring complex reasoning, was the addition of a series of intermediate reasoning steps, i.e., chain-of-thought (CoT) prompting [24, 53]. CoT was incorporated in many LLMs, being executed by default in the so-called reasoning models [54]. Another approach to adapt foundational models for more specific tasks has been the use of agents, i.e., different instances of LLMs responsible for specific tasks, that are orchestrated to reach a common goal. Agentic AI became a very active area of research [15, 55, 56].

Generative capabilities now extend far beyond text. High-fidelity images, videos, and design prototypes can now be generated from textual descriptions [29, 57]. GenAI systems now generate multi-modal content, which is particularly relevant for IS and organizations which generate diverse data formats. The multi-modal systems generally have similar design principles, and the large transformer networks are trained on massive paired datasets (e.g., text–speech, text–image), so that the trained embedding spaces allow for mapping from one modality into another.

Besides transformers, other key GenAI technologies are Generative Adversarial Networks (GANs), Variational Autoencoders (VAEs), and Diffusion Models. GANs consist of two neural networks, a generative, and an adversarial, which compete with each other to produce new content. VAEs produce encodings, i.e., a compressed latent space, and add variations to that space to generate new content. Diffusion models gradually add noise to training data, and then learn reversely to learn the original data gradually. Although the learning idea resembles an encoding-decoding process, the gradual learning process provides a new quality to diffusion models.

GenAI is not only a major technological breakthrough, but also a socio-technical organizational infrastructure [9, 12] that shapes the future of work [5, 11, 14, 36], decision-making [13], and digital platforms. [35] argue that GenAI's conversational interfaces, open-ended generativity, and human-like communication capabilities create new forms of human–AI interaction and hybrid intelligence. Similarly, research on decision-support systems highlights the potential of GenAI to synthesize information, generate alternative scenarios, and serve as an interactive interface to organizational analytics and databases [15, 29, 58]. At the ecosystem level, GenAI is reshaping digital platforms by enabling AI-augmented services, new forms of value creation, and changes in competitive dynamics [3].

GenAI is a challenging technology. Their probabilistic nature, and the implications of this, are sometimes not easy to grasp [29, 59]. Additionally, there are well-known challenges identified concerning fairness, transparency, provenance, and model explainability, which are also characteristics of other machine-learning algorithms. Therefore, the integration of GenAI models into organizations requires new frameworks for responsible AI, and human oversight [9, 15, 28, 29, 32].

In general, GenAI represents a fundamental shift in information production and user-machine interaction. The combination of a general-purpose decision support system tool, with multi-modal content and easy integration into organizational processes, makes it already an indispensable technology for IS research related to digital transformation, organizational decision-making, human–machine collaboration, and responsible technology management.

### 2.2. Information systems perspective

While Information Systems (IS) and Software Engineering (SE) share a common interest in the design and development of Information Technology (IT) for human use, IS research is distinguished by its socio-technical emphasis. In IS, information systems are viewed not only as technical systems but also as social and organizational ones, comprising technical, organizational, and semiotic components (e.g., [60–63]).

The socio-technical approach has been widely discussed in IS already during 1970s and 1980s, with ETHCIS and SSM methodologies as examples advocating it. The socio-technical approach argues for the joint optimization of both social and technical components, recognizing their interaction [64]. Within such a framing, IS as a discipline is broadly interested in the application of IT artifacts to support particular task(s) embedded within particular context(s), with an aim to increase our understandings of IS design, use and impacts, i.e., of "(1) how IT artifacts are conceived, constructed, and implemented, (2) how IT artifacts are used, supported, and evolved, and (3) how IT artifacts impact (and are impacted by) the contexts in which they are embedded" [65].

As [65] indicate, IS research is interested in a variety of aspects intertwined with the design, use and impacts of IT. IS research has ranged from individual to organizational and society level analyses, including studies, for example, on individual technology acceptance [66], organizational digital transformation [67, 68], and societal level implications of IT [69].

The discipline, with its socio-technical emphasis, has since its early days had an interest in power, politics, and ethics as intermingled with IT (e.g., [64, 70, 71]). Recently, there has been an increased interest to address such concerns (e.g., [63, 72]), especially in the context of emerging technologies, including AI, and their regulation (e.g,. [9, 32, 69, 73, 74]). Moreover, while the original focus of IS research has been on work, organizations, business and management (e.g., [60–63]), during the past decades, it has been acknowledged that IT has spread to different everyday contexts and life spheres beyond work and organizations [75], with IS researchers nowadays addressing various user groups, usages, and impacts of IT in diverse everyday contexts, including, for example, studies with children, people with special needs or marginalized communities (e.g., [72, 76–78]).

Overall, IS as a discipline addresses a broader range of concerns compared to SE, which is more technology and engineering focused discipline, even if also SE has acknowledged the significance of human and social factors for SE practice already for long (e.g., [79, 80]), and also recently related to GenAI [23, 81].

## 3. Related work

For this study, we consider SLRs, tertiary studies, other literature surveys, and also research agendas or roadmaps on GenAI from *other* computing disciplines as the related work. As our SLR was focused specifically on IS literature, such papers from related computing disciplines were not included in our SLR, whereas related IS research agendas and secondary studies are already covered as a part of our SLR results. Thus, in this section, we discuss related work not covered by our SLR. Primarily, we discuss related work from the field of software engineering (SE) specifically.

As there is a plethora of related work that has been published in related disciplines, we aim to give a general overview of it, acknowledging some of the more directly related work that has been published outside IS literature. We therefore focus on related work discussing GenAI or LLMs more generally, as opposed to work with more specific scopes related to GenAI (e.g., GenAI specifically for software testing). Moreover, we limit this discussion to peer-reviewed works, thus excluding the numerous pre-prints on the topic.

*First*, the related work includes SLRs and other literature surveys from related disciplines. In terms of more general SLRs, [82] and [83] both review studies on using LLMs for SE overall, and were published in 2025 and 2024 respectively. While [84] discuss GenAI for "optimising the software engineering process", their SLR is similarly wide in scope, covering various SE use cases. [85] also conduct a survey of literature on LLMs for SE overall. [86] conduct an MLR on the role of GenAI in SE overall. Finally, more specific but still quite general, [87] conduct on SLR on GenAI in Agile SE.

Past such more general literature reviews and surveys, we are able to identify various reviews and surveys with more specific foci. These include the following areas, among others; this is by no means an exhaustive list or survey, as such studies are numerous. *Security and privacy:* [88], [89], [90], and [91] all review literature on LLMs in relation to security, such as code security or cybersecurity overall. *Software testing:* [92] and [93] survey literature related to software testing with LLMs, in addition to a number of literature reviews on using AI overall for software testing. *Requirements engineering:* both [94] and [95] conduct SLRs on GenAI for requirements engineering. These are but three examples of some of the more specific foci seen in literature studies from related disciplines, among many others, as providing a systematic review of such studies is out of the scope of this paper.

*Second*, we consider research agendas or roadmaps from related disciplines as related work. [96] present a research agenda for GenAI for SE overall. While the research roadmap of [97] discusses AI overall for SE rather than GenAI specifically, much of their roadmap is nonetheless related to recent advances in GenAI and LLMs. While we were able to identify various research agenda papers discussing GenAI as a part of the agenda otherwise focused on some other topic, we have only included papers more focused on GenAI specifically here.

In Table 1, we provide an overview of related work from other computing disciplines with a brief comparison to our work. The table includes only related work with a general viewpoint. This means that it includes papers discussing "GenAI in SE", but not papers focused on more specific application contexts like "GenAI in requirements engineering (in SE)", of which there are currently plenty as we have highlighted earlier in this section.

In brief, we were unable to identify a notable number of research roadmaps and research agendas with a more general viewpoint like ours. We identified two such SE papers. However, neither of these were supplemented by a systematic review of literature like ours. While we have included in the table some SLRs focused on *primary* studies, we were unable to identify tertiary studies focusing on GenAI like ours for the time being.

## 4. Method

The rapidly expanding body of research on GenAI presents multiple avenues for evidence synthesis, ranging from aggregating individual primary studies to conducting higher-level reviews that integrate consolidated knowledge. While synthesizing primary studies provides granular insights into specific phenomena, the GenAI-in-IS literature has evolved in a way that disperses contributions across heterogeneous empirical designs, disciplinary contexts, and emerging conceptual framings. In contrast, the field has begun to produce secondary studies and research agenda papers that already distill key themes and articulate future directions.

### 4.1. Methodological framework

Reviewing this specific layer of literature – comprising both retrospective syntheses and prospective agendas – necessitates a methodological strategy that extends beyond standard protocols typically designed for synthesizing homogeneous sets of primary studies.

To address this challenge, we adopted a tailored methodological approach informed by established guidelines from multiple complementary sources. Rather than rigidly adhering to a single SLR standard, we integrated the foundational SLR guidelines by [98] for structuring the review protocol with the practical evidence synthesis procedures from [99] to guide our search and screening strategy. To ensure disciplinary alignment with Information Systems, particularly regarding our qualitative synthesis, we incorporated the methodological guidance by [100], while simultaneously consulting recent recommendations for software engineering secondary studies by [101] to address the specific methodological requirements of tertiary-level synthesis. Additionally, we adopted the framework by [102] to structure our Threats to Validity section. This multi-perspective approach allowed us to maintain methodological rigor while flexibly accommodating the distinctive requirements of synthesizing both secondary studies and research agenda papers within the Information Systems domain.

Table 1. Comparison of our study to related work identified from other computing disciplines.

| Authors | Title | RQs or ROs | Difference compared to our RQs |
|---|---|---|---|
| Zheng et al. | Towards an understanding of large language models in software engineering tasks | RQ1: What are the current works focusing on combining LLMs and software engineering?<br>RQ2: Can LLMs truly help better perform current software engineering tasks? | Focus on SE domain, performance metrics. Not a research roadmap or agenda. |
| Hou et al. | Large language models for software engineering: A systematic literature review | RQ1: What LLMs Have Been Employed to Date to Solve SE Tasks?<br>RQ2: How Are SE-Related Datasets Collected, Pre-Processed, and Used in LLMs?<br>RQ3: What Techniques Are Used to Optimize and Evaluate LLM4SE?<br>RQ4: What SE Tasks Have Been Effectively Addressed to Date Using LLM4SE? | Focus on SE domain, SE tasks, datasets, and LLM optimizatio and evalution techniques for SE. Not a research roadmap or agenda. |
| Karlovs | Generative Artificial Intelligence Use in Optimising Software Engineering Process: A Systematic Literature Review | RQ1: Which Software Engineering sub-fields are actively experimented on with Generative AI and which ones are underrepresented?<br>RQ2: Who are the active researchers in the field for future collaboration?<br>RQ3: What are the most common research methods used in the field of Generative AI for Software Engineering research? | Focus on SE domain, focus on methodology. Not a research roadmap or agenda. |
| Fan et al. | Large language models for software engineering: Survey and open problems | No RQs/ROs explicitly defined. Implicit RO statement: *"This paper surveys the recent developments, advances and empirical results on LLM-based SE; the application of Large Language Models (LLMs) to Software Engineering (SE) applications."* | Focus on SE domain, SE use cases. Not a research roadmap or agenda. |
| Bazzan et al. | Analysing the Role of Generative AI in Software Engineering-Results from an MLR | RQ1: What is Generative AI? RQ2: How is Generative AI used in Software Engineering? RQ3: What are the benefits associated with using Generative AI in Software Engineering? RQ4: What are the risks associated with using Generative AI in Software Engineering? | Focus on SE domain. RQ3 and 4 very similar to our RQ2, but review focuses on primary studies and grey literature. Not a research roadmap or agenda. |
| Cornide-Reyes et al. | Generative Artificial Intelligence in Agile Software Development Processes: A Literature Review Focused on User eXperience | RQ1: How can GenAI tools optimize user experience in agile software development projects? RQ2: What are agile teams' main challenges when integrating GenAI tools into software development projects? and RQ3: What stages of the agile software development cycle benefit from implementing GenAI tools? | Focus specifically on agile software development (ISD or SE). Not a research roadmap or agenda. |
| Nguyen-Duc et al. | Generative artificial intelligence for software engineering – A research agenda | No RQs/ROs explicitly defined. Implicit RO statement formulated as challenge/gap: *"we do not have an overall picture of the current state of GenAI technology in practical software engineering usage scenarios."* | Difference: Focus on SE domain, practical SE use cases. No SLR conducted as part of the research agenda. |
| Ahmed et al. | Artificial Intelligence for Software Engineering: The Journey so far and the Road ahead | No RQs/ROs explicitly defined. Implicit RO statement: [to] *"highlight the recent deep impact of artificial intelligence on software engineering by discussing successful stories of applications of artificial intelligence to classic and new software development challenges".* | Focus on SE domain. No SLR conducted as part of the research roadmap. |
| Kotti et al. | Machine learning for software engineering: A tertiary study | No RQs/ROs explicitly defined. Contributions formulated as follows: *"we identify what SE tasks have been tack-led with ML techniques, which SE knowledge areas could be better covered by ML techniques aswell as the prominent ML techniques applied in SE. We also provide a classification scheme for cat-egorizing ML techniques in SE along four axes."* | Focus on SE domain. Tertiary study but focused on AI/ML overall, not GenAI specifically. |

#### 4.1.1. Rationale for hybrid synthesis design

This study adopts a hybrid synthesis design that integrates two methodologically distinct literature types: (1) secondary studies that retrospectively synthesize empirical evidence from primary research, and (2) research agenda papers that prospectively articulate expert-driven propositions about future directions. We explicitly justify this design choice and explain how it enhances rather than compromises methodological rigor.

The rationale for combining these literature types stems from the nascent and rapidly evolving nature of the GenAI field. As noted by [101], mixed-methods approaches are "particularly important for industry-based interventions, when outcomes are influenced by the complex nature of the relationship between the intervention and its environment." GenAI in IS represents precisely such a complex intervention: its impacts span technical, organizational, and societal dimensions, and its rapid evolution means that empirical evidence inevitably lags behind practitioner experience and expert foresight.

Secondary studies (such as SLRs and scoping reviews) provide retrospective synthesis of what empirical research has established about GenAI's benefits, challenges, and applications. However, in a field where the foundational technology (transformer-based Large Language Models) and tools like ChatGPT have only been widely accessible since late 2022, such retrospective evidence necessarily captures only a narrow temporal window and may not reflect emerging concerns or opportunities. Research agenda papers, authored by domain experts, provide prospective analysis of where the field should direct its attention. These papers synthesize expert judgment, identify gaps in current knowledge, and propose directions that may not yet have accumulated sufficient empirical evidence for inclusion in traditional secondary studies.

By synthesizing both types of literature, this study provides a more complete picture than either source alone could offer. The retrospective evidence from secondary studies grounds our findings in empirical reality, while the prospective propositions from research agendas extend our analysis to emerging concerns and future-oriented recommendations. This approach aligns with the fourth step of Evidence-Based Software Engineering (EBSE), which concerns "integrating the critical appraisal with software engineering expertise and stakeholders' values" [103].

Importantly, we maintain methodological transparency by clearly distinguishing these two evidence streams throughout our analysis. Tables 4 and 5 separately catalogue secondary studies (coded S01–S18) and research agenda papers (coded R01–R10). In our synthesis, we attribute findings to their source type, enabling readers to assess the evidentiary basis for each synthesized theme. This approach is consistent with the recommendation by [104] that "it is critical that readers know the provenance of all recommendations, so they can properly judge their credibility."

#### 4.1.2. Differentiated treatment of literature types

While our hybrid synthesis integrates findings from both secondary studies and research agenda papers, we acknowledge that these literature types have fundamentally different epistemological characteristics, which necessitate differentiated treatment in quality assessment and synthesis.

**Epistemological distinctions.** Secondary studies (S01–S18) represent systematic aggregations of primary empirical research. Their value derives from the rigor of their review methodology and the quality of the primary studies they synthesize. As such, they can be assessed using established criteria for evaluating systematic reviews, such as the DARE (Database of Abstracts of Reviews of Effects) criteria [105]. Secondary studies report findings that are, in principle, traceable to underlying primary evidence.

Research agenda papers (R01–R10) represent expert-driven conceptualizations of future research needs. Their value derives from the expertise and insight of their authors, the comprehensiveness of their environmental scanning (which involves monitoring and analyzing the broader external context – technological, social, and industrial trends – to identify emerging threats and opportunities that should guide future research directions), and the actionability of their proposed directions. These papers do not claim to synthesize empirical evidence in the same manner as secondary studies; rather, they offer informed scholarly judgment about where research attention should be directed. As [104] note, expert opinion can provide valuable insights but should be distinguished from empirically-grounded evidence.

**Synthesis strategy.** Our synthesis employs a *parallel-but-integrated* approach. During thematic analysis, we coded both literature types using the same coding scheme to enable the identification of common themes. However, we maintain attribution to source type throughout, allowing readers to distinguish empirically-grounded findings (derived primarily from secondary studies) from expert-proposed directions (derived primarily from research agendas).

In the Discussion section, we analyze findings across these two evidence streams. Where secondary studies and research agendas converge on the same themes, this convergence strengthens confidence in those findings. Where they diverge – for example, where research agendas propose directions not yet reflected in secondary study findings – this may indicate emerging areas that have not yet attracted sufficient empirical attention. Conversely, where secondary studies report challenges that research agendas do not identify, this may indicate blind spots in current expert discourse.

This approach is consistent with mixed-methods review methodology as described by [106]: qualitative and quantitative findings are synthesized separately, then compared and integrated to produce more nuanced overall findings. In our case, we adapt this approach to integrate retrospective evidence synthesis with prospective expert judgment.

**Implications for interpretation.** Readers should interpret our synthesized findings with awareness of their evidentiary basis. Findings supported primarily by secondary studies (e.g., many of the challenges in categories C1–C4) have stronger empirical grounding but may reflect a narrower temporal window. Findings supported primarily by research agendas (e.g., some of the future directions in categories F1–F6) represent expert consensus about important directions but await empirical validation. Findings supported by both types of literature represent the strongest form of convergent evidence available in this hybrid synthesis.

#### 4.1.3. Reproducibility and tool support

To support transparency, reproducibility, and methodological scrutiny, we provide a complete replication package – including search logs, screening decisions, extraction sheets, and coding scripts – on GitHub[1]. Several tools supported the process: `Zotero` for reference management and PDF annotation, `Rayyan` for the collaborative screening of bibliographic records, `Google Sheets` for structured data extraction, `Python` for documentation and audit trail analysis, and `PowerPoint` for thematic visualizations.

[1] https://github.com/przybylek/GenAI4IS

### 4.2. Research questions

The review was guided by three research questions (RQs), each targeting a distinct dimension of the current knowledge landscape on GenAI in IS:

**RQ1:** What is the landscape of secondary studies and research agenda papers on GenAI in Information Systems?
- **RQ1.1:** What types of papers have been published (e.g., SLR, mapping study, research agenda)?
- **RQ1.2:** What are the publication trends over time and across venues?
- **RQ1.3:** What specific application sectors are covered?

**RQ2:** What benefits and challenges have been identified in the use of Generative AI within Information Systems?

**RQ3:** What research gaps and future research directions have been proposed?

**RQ1** examines the structure of the evidence base – types of studies, publication trends, and application sectors. **RQ2** synthesizes the reported benefits and challenges associated with the use of GenAI in IS. **RQ3** aggregates reported research gaps and recommended research directions. Together, these questions enable a comprehensive view of the present state of knowledge and the projected research trajectory in this rapidly evolving domain.

### 4.3. Search strategy

The search strategy was designed to identify relevant secondary studies and research agenda papers on GenAI in IS. We targeted three bibliographic repositories that provide broad and authoritative coverage of IS research: Scopus[2], the Web of Science Core Collection (WoS)[3], and the AIS eLibrary (AISeL)[4].

#### 4.3.1. Search string construction

Following established guidelines [107], we constructed a search string composed of three distinct fragments, joined by the **AND** operator:
- **Study Type:** Targets terms identifying secondary studies (e.g., systematic review or mapping study) or forward-looking papers (e.g., "research agenda" or "roadmap").
- **Phenomenon:** Includes keywords for the core topic of interest – Generative AI – using both broad terms (e.g., "generative AI") and specific, highly prevalent examples (e.g., "ChatGPT") to maximize coverage.
- **Domain:** Scopes the search using terms that characterize the IS discipline or closely related areas.

The generic search string is as follows:

```
(SLR OR "systematic review" OR "systematic mapping" OR "mapping study" OR
↪  "literature survey" OR "literature review" OR "scoping review" OR
↪  "meta-analysis" OR "tertiary study" OR "secondary study" OR "research
↪  agenda" OR "roadmap")
AND
```

[2] https://www.scopus.com/
[3] https://www.webofknowledge.com
[4] https://aisel.aisnet.org/

```
("large language model" OR "LLM" OR "generative AI" OR "GenAI" OR "Gen AI"
↪  OR "generative artificial intelligence" OR "ChatGPT" OR "GPT" OR
↪  "conversational AI" OR "artificial intelligence language model" OR "AI
↪  language model")
AND
("information systems" OR "MIS" OR "information technology" OR
↪  "informatics" OR "project management")
```

#### 4.3.2. Search execution

The search string was adapted to the syntax and field constraints of each database while preserving its semantic integrity. We limited the search to publications from 2023 onward and executed it on April 12, 2025. This temporal scope was intentionally selected to capture the accelerated surge of GenAI-related research that followed the public release of ChatGPT in late 2022, ensuring that our review reflects the period in which substantive scholarly engagement with GenAI began to emerge.

Table 2 presents the number of records returned from each database and the specific fields that were queried. The initial search yielded a total of 242 documents before duplicate removal.

Table 2. Database search strategy and initial results

| Database | Fields Searched | Results |
|---|---|---|
| Scopus | TITLE-ABS-KEY | 105 |
| AIS eLibrary (AISeL) | All Metadata Fields | 102 |
| Web of Science (WoS) | Title, Abstract, Author Keywords | 35 |
| Total initial records | | 242 |

### 4.4. Inclusion and exclusion criteria

To systematically filter the studies identified during the search, we established a set of inclusion and exclusion criteria. These criteria were formulated based on our research questions to ensure that only the most relevant studies were retained for data extraction and synthesis.

The inclusion criteria (IC) define the necessary attributes for a paper to be included in our review, while the exclusion criteria (EC) specify conditions that lead to a paper's exclusion. A study was carried forward to the analysis phase only if it met all inclusion criteria and met none of the exclusion criteria. The criteria are detailed in Table 3.

### 4.5. Study selection

The study selection process followed a multi-stage filtering approach to systematically reduce the initial set of 242 documents to the final set of relevant studies. The entire process is visually summarized in the PRISMA flow diagram in Figure 1.

All bibliographic records were exported in RIS format from the three databases and imported into `Rayyan.ai`, which automatically detected and flagged duplicates. After removing duplicates, 210 unique records remained for screening.

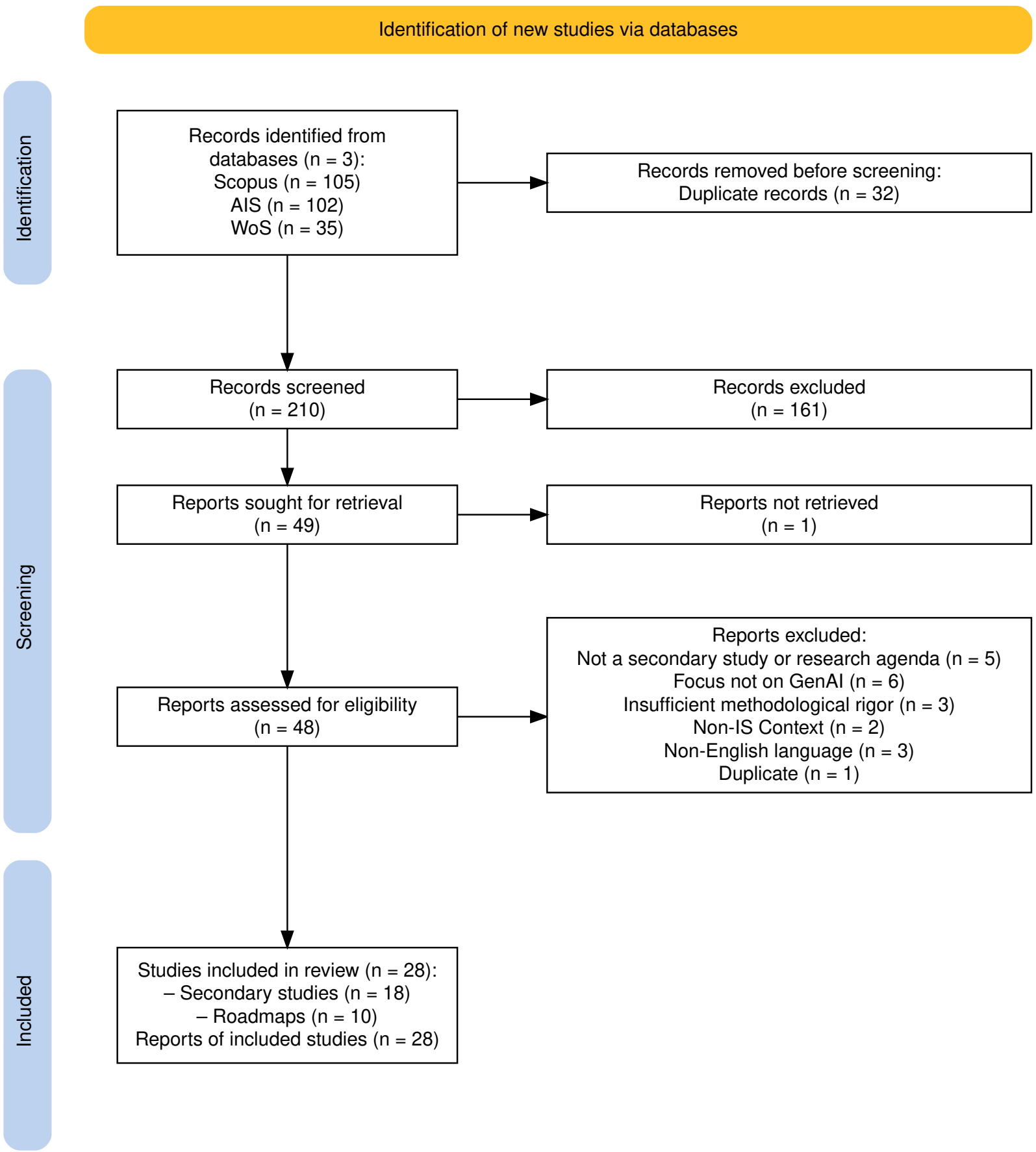

Figure 1. PRISMA flow diagram [108]

#### 4.5.1. Phase 1: Title and abstract screening

Three researchers independently screened the titles and abstracts of the 210 records using the predefined inclusion and exclusion criteria (Table 3). Prior to screening, a calibration meeting (17 April 2025) was held to ensure consistent interpretation of the criteria.

Table 3. Inclusion and exclusion criteria

| ID | Inclusion criteria |
|---|---|
| IC1 | The paper is classified as a secondary study or a research agenda/roadmap paper. |
| IC2 | The paper's primary focus is on Generative AI; a superficial or passing mention is not sufficient. |
| IC3 | The study is situated within the context of Information Systems or a closely related domain (such as Information Technology or Project Management). |
| IC4 | The paper is a peer-reviewed publication (e.g., journal article, conference paper). |
| IC5 | The paper provides sufficient methodological detail to allow for an assessment of its rigor (e.g., describes the search process or analysis method). |
| IC6 | The paper is written in English. |
| **ID** | **Exclusion criteria** |
| EC1 | The paper is a purely bibliometric or scientometric analysis without a qualitative synthesis of the literature's contributions. |
| EC2 | The paper is written in a language other than English. |
| EC3 | The full text of the paper could not be retrieved through institutional subscriptions or publicly available archives. |
| EC4 | The paper is a duplicate of another study already included in the review. |

Independent assessments produced an initial inter-rater agreement of 74%. All conflicts (54 cases) and papers marked "maybe" by one or more reviewers (33 cases) were resolved in a consensus meeting on 15 May 2025, occasionally supported by brief full-text inspection. This phase resulted in the exclusion of 161 papers and the advancement of 49 studies to the next phase. The list of these studies is available in the online appendix[5].

#### 4.5.2. Phase 2: Integrated full-text screening and data extraction

To improve both rigor and efficiency, we combined full-text screening with data extraction into a single integrated phase. This ensured that inclusion decisions were made only after a complete and detailed reading of each study, reducing the likelihood of erroneous exclusions and producing higher-quality extracted data.

**Pilot phase and protocol refinement.** The phase began with a pilot in which the three core researchers independently applied a draft extraction protocol to three common papers. Insights from the pilot were used to refine the extraction form, clarify decision rules, and establish a shared approach to evaluating borderline cases. A consensus meeting was then used to finalize the protocol, after which three additional researchers were trained, bringing the total team to six reviewers.

**Execution and quality assurance.** Before screening, all full texts were verified in Zotero. Four papers were excluded at this stage: one previously undetected duplicate, two non-English papers, and one paper with an inaccessible full text (despite attempts to contact the authors). The remaining 42 studies were distributed across the six reviewers, who conducted full-text reading while simultaneously extracting data into the shared Google Sheet and annotating relevant PDF passages. Direct quotations were recorded verbatim to maintain traceability.

To ensure consistent application of the protocol, the core researchers hosted eight weekly alignment meetings where reviewers could raise questions and discuss ambiguous cases. A quality audit revealed deficiencies in one reviewer's extraction work; the reviewer subsequently withdrew, and their assigned papers were re-evaluated by a new (trained) team member.

Following the integrated screening and extraction, 17 papers were excluded for failing the inclusion criteria upon full-text inspection. This resulted in a final set of 28 studies: 18 secondary studies and 10 research agenda papers. Each included study was assigned a unique identifier – "S#" for secondary studies and "R#" for agenda papers – as listed in Table 4 and Table 5, respectively. The entire selection phase spanned approximately seven weeks and concluded on 16 July 2025.

### 4.6. Data extraction

As described in Section 4.5.2, data extraction was conducted concurrently with full-text screening during Phase 2. Data was extracted using a structured form implemented in Google Sheets. This form was developed based on our RQs and iteratively refined during the pilot phase. Table 6 details the data fields, their descriptions, and the research questions they primarily address.

[5] https://github.com/przybylek/GenAI4IS

Table 4. List of included secondary studies (S studies).

| ID | Authors | Title | Venue type | Venue | WoS | Year |
|---|---|---|---|---|---|---|
| S01 | Meng, X. et al. | The application of large language models in medicine: a scoping review | Journal | iScience | Q1 | 2023 |
| S02 | Beheshti, M. et al. | Evaluating the reliability of ChatGPT for health-related questions: a systematic review | Journal | Informatics | Q3 | 2025 |
| S03 | Bellanda, V.C.F. et al. | Applications of ChatGPT in the diagnosis, management, education, and research of retinal diseases: a scoping review | Journal | International Journal of Retina and Vitreous | Q2 | 2024 |
| S04 | Bendig, D. and Bräunche, A. | The role of artificial intelligence algorithms in information systems research: a conceptual overview and avenues for research | Journal | Management Review Quarterly | Q1 | 2024 |
| S05 | Bracken, A. et al. | Artificial Intelligence (AI)–powered documentation systems in healthcare: a systematic review | Journal | Journal of Medical Systems | Q1 | 2025 |
| S06 | Clear, T. et al. | AI integration in the IT professional workplace: a scoping review and interview study with implications for education and professional competencies | Conference | Innovation and Technology in Computer Science Education | NA | 2025 |
| S07 | Ghebrehiwet, I. et al. | Revolutionizing personalized medicine with generative AI: a systematic review | Journal | Artificial Intelligence Review | Q1 | 2024 |
| S08 | Gumusel, E. | A literature review of user privacy concerns in conversational chatbots: a social informatics approach: an annual review of information science and technology (ARIST) paper | Journal | Journal of the Association for Information Science and Technology | Q1 | 2025 |
| S09 | Laine, J. et al. | Understanding the ethics of Generative AI: established and new ethical principles | Journal | Communications of the Association for Information Systems | Q3 | 2025 |
| S10 | Lareyre, F. et al. | Comprehensive review of natural language processing (NLP) in vascular surgery | Journal | EJVES Vascular Forum | Q2 | 2023 |
| S11 | Li, M. and Guenier, A.W. | ChatGPT and health communication: a systematic literature review | Journal | International Journal of E-Health and Medical Communications | Q4 | 2024 |
| S12 | Maita, I. et al. | Pros and cons of Artificial Intelligence-ChatGPT adoption in education settings: a literature review and future research agendas | Journal | IEEE Engineering Management Review | — | 2024 |
| S13 | Mambile, C., and Ishengoma, F. | Exploring the non-linear trajectories of technology adoption in the digital age | Journal | Technological Sustainability | — | 2024 |
| S14 | Mohammad, B. et al. | The pros and cons of using ChatGPT in medical education: a scoping review | Conference | Healthcare Transformation with Informatics and Artificial Intelligence | NA | 2023 |
| S15 | Onatayo, D. et al. | Generative AI applications in architecture, engineering, and construction: trends, implications for practice, education and imperatives for upskilling – a review | Journal | Architecture | Q3 | 2024 |
| S16 | Ouanes, K. | Generative artificial intelligence in healthcare: current status and future directions | Journal | Italian Journal of Medicine | Q4 | 2024 |
| S17 | Pool, J. et al. | Large language models and generative AI in telehealth: a responsible use lens | Journal | Journal of the American Medical Informatics Association | Q1 | 2024 |
| S18 | Schneider, J. | Explainable Generative AI (GENXAI): a survey, conceptualization, and research agenda | Journal | Artificial Intelligence Review | Q1 | 2024 |

Table 5. List of included roadmaps (R studies)

| ID | Authors | Title | Venue type | Venue | WoS | Year |
|---|---|---|---|---|---|---|
| R01 | Wei, X. et al. | Addressing bias in Generative AI: challenges and research opportunities in information management | Journal | Information and Management | Q1 | 2024 |
| R02 | Chau, M. and Xu, J. | An IS research agenda on Large Language Models: development, applications, and impacts on business and management | Journal | ACM Transactions on Management Information Systems | Q2 | 2025 |
| R03 | Dwivedi, Y.K. et al. | GenAI's impact on global IT management: a multi-expert perspective and research agenda | Journal | Journal of Global Information Technology Management | Q2 | 2025 |
| R04 | Feuerriegel, S. et al. | Generative AI | Journal | Business and Information Systems Engineering | Q1 | 2024 |
| R05 | Haase, J. et al. | Interdisciplinary directions for researching the effects of robotic process automation and large language models on business processes | Journal | Communications of the Association for Information Systems | Q3 | 2024 |
| R06 | Jarvenpaa, S. and Klein, S. | New frontiers in information systems theorizing: human-gAI collaboration | Journal | Journal of the Association for Information Systems | Q1 | 2024 |
| R07 | Nah, F.F.H. et al. | An activity system-based perspective of generative AI: challenges and research directions | Journal | AIS Transactions on Human-Computer Interaction | — | 2023 |
| R08 | Sigala, M. et al. | ChatGPT and service: opportunities, challenges, and research directions | Journal | Journal of Service Theory and Practice | Q2 | 2024 |
| R09 | Srivastava, A. et al. | The present and future of AI: ethical issues and research opportunities | Journal | Communications of the Association for Information Systems | Q3 | 2025 |
| R10 | Storey, V.C. et al. | Generative Artificial Intelligence: Evolving technology, growing societal impact, and opportunities for information systems research | Journal | Information Systems Frontiers | Q1 | 2025 |

Table 6. Data extraction form

| Data Field | Description | Mapped RQ(s) |
|---|---|---|
| | Bibliographic and contextual data | |
| Type of publication | Classification as either “Journal” or “Conference”. | RQ1.2 |
| IS Application sectors | The industry or organizational contexts in which GenAI applications are examined (e.g., healthcare, education, manufacturing). | RQ1.3 |
| | Methodological details – secondary studies | |
| Type of Secondary Study | The specific review method used (e.g., SLR, SMS). | RQ1.1 |
| Databases searched | List of electronic databases used to find primary literature. | RQ1 |
| Number of primary studies included | The total count of primary studies included in the review. | RQ1 |
| Publication date range | The publication date range of the primary studies included. | RQ1 |
| | Methodological details – research agendas | |
| Research method used | The research method used to identify research gaps and/or directions. | RQ1.1 |
| | Data for Qualitative Synthesis | |
| Reported benefits | Quoted or summarized benefits, advantages, or opportunities. | RQ2 |
| Reported challenges or limitations | Quoted or summarized challenges, risks, ethical concerns, or limitations. | RQ2 |
| Suggestions about gaps or future research | Identified research gaps and specific suggestions for future work. | RQ3 |

Table 7. Quality assessment of secondary studies

| Study | C1 | C2 | C3 | C4 | C5 | Score |
|---|---|---|---|---|---|---|
| S01 | Yes | Yes | Partly | Partly | Yes | 0.8 |
| S02 | Yes | Partly | Partly | Yes | Yes | 0.8 |
| S03 | Yes | Yes | Yes | Yes | Yes | 1.0 |
| S04 | Partly | Yes | No | Yes | No | 0.5 |
| S05 | Yes | Yes | Yes | Yes | Partly | 0.9 |
| S06 | Yes | Partly | No | No | Partly | 0.4 |
| S07 | Yes | Partly | Partly | Yes | Yes | 0.8 |
| S08 | Yes | Yes | Yes | Yes | Yes | 1.0 |
| S09 | Yes | Yes | No | No | Yes | 0.6 |
| S10 | Partly | No | No | No | Partly | 0.2 |
| S11 | Yes | Yes | Yes | Yes | Yes | 1.0 |
| S12 | Yes | No | No | Partly | Partly | 0.4 |
| S13 | Partly | Yes | No | Yes | Yes | 0.7 |
| S14 | No | Yes | No | Yes | Partly | 0.5 |
| S15 | No | Yes | No | Yes | Yes | 0.6 |
| S16 | No | No | No | No | No | 0.0 |
| S17 | Yes | Yes | No | Yes | Yes | 0.8 |
| S18 | No | No | No | Yes | Yes | 0.4 |

### 4.7. Quality assessment

Following the SEGRESS guidelines [101], we conducted quality assessment of all included studies. Given the hybrid nature of our synthesis – combining secondary studies with research agenda papers – we applied differentiated assessment criteria appropriate to each literature type.

#### 4.7.1. Assessment of secondary studies

For the 18 secondary studies (S01–S18), we applied the DARE criteria, which are specifically designed for assessing the methodological quality of systematic reviews [105]. The five DARE criteria are as follows:
1. Are the review's inclusion and exclusion criteria described and appropriate?
2. Is the literature search likely to have covered all relevant studies?
3. Did the reviewers assess the quality/validity of the included studies?
4. Were basic data/studies adequately described?
5. Were the included studies synthesised?

Each criterion was rated as *Yes* (1.0), *Partly* (0.5), or *No* (0.0). A composite quality score was calculated for each study as the arithmetic mean across the five criteria. The resulting scores and detailed ratings are reported in Section 4.7.3.

#### 4.7.2. Assessment of research agenda papers

The DARE criteria are designed for systematic reviews and are not directly applicable to research agenda papers, which represent expert-driven conceptualizations rather than systematic evidence syntheses. Following the principle of differentiated treatment outlined in Section 4.1.2, we developed custom quality criteria appropriate to this literature type:
1. Is the method for identifying research gaps or directions transparently described?
2. Are the proposed directions grounded in cited empirical evidence or prior reviews?
3. Are the proposed research directions actionable?
4. Does the paper consider more than one stakeholder perspective (e.g., technical, organizational, societal)?
5. Is the scope and context of applicability clearly delimited?

These criteria assess the transparency, empirical grounding, actionability, stakeholder comprehensiveness, and delimitation of research agenda papers. The same rating scale (0.0–1.0) was applied.

#### 4.7.3. Quality assessment results

Each of the 28 studies was independently assessed by two reviewers using the defined criteria (140 rating items in total). Observed agreement was 69.3% (97/140 items), yielding Cohen's $\kappa = 0.47$ ($z = 7.01$, $p < 0.001$), indicating moderate agreement. All disagreements were subsequently resolved through discussion until consensus was reached.

**Secondary studies results.** Of the 18 secondary studies, three (S03, S08, S11) met all five DARE criteria, achieving a perfect score of 1.0. The most prevalent limitation concerned criterion 3, as many studies failed to formally assess the quality of the primary literature they synthesized. Table 7 presents the detailed ratings.

**Research agenda results.** Among the 10 research agenda papers, one (R09) met all five custom criteria. The most common limitation was criterion 1 (methodological transparency), reflecting the expert-opinion nature of this document class. Table 8 presents the detailed ratings.

Table 8. Quality assessment of research agenda papers

| Study | C1 | C2 | C3 | C4 | C5 | Score |
|---|---|---|---|---|---|---|
| R01 | No | Partly | Yes | Yes | Yes | 0.7 |
| R02 | No | Yes | Yes | Partly | Yes | 0.7 |
| R03 | Partly | Yes | Partly | Yes | Partly | 0.7 |
| R04 | Partly | Partly | Partly | Yes | Partly | 0.6 |
| R05 | Partly | Yes | Partly | Yes | Yes | 0.8 |
| R06 | Partly | Partly | Yes | No | Partly | 0.5 |
| R07 | Partly | Yes | Yes | Yes | Yes | 0.9 |
| R08 | Partly | Yes | Yes | Yes | Yes | 0.9 |
| R09 | Yes | Yes | Yes | Yes | Yes | 1.0 |
| R10 | Partly | Yes | Partly | Yes | Partly | 0.7 |

**Inclusion strategy.** Importantly, no study was excluded on the basis of its quality score. Because GenAI in the IS field has only produced a limited corpus of secondary studies and research agendas since late 2022, excluding lower-scoring contributions would narrow an already small evidence base and risk suppressing early yet potentially valuable insights.

#### 4.7.4. Certainty of evidence

To enable readers to assess confidence in each synthesized finding, we adopted a *structured transparency* approach. Rather than imposing a single assessor-driven confidence rating – which would introduce an additional layer of subjective interpretation – we make the evidential basis of each finding directly inspectable.

For every synthesized code (the most granular unit of our thematic structure), we report the complete set of contributing studies together with their individual quality scores. This design allows readers to immediately see how many studies support a given finding and the methodological rigor of those studies. This information is presented in Tables 9, 10, and 11 (Section 5.2).

### 4.8. Data analysis and synthesis

The analysis of the extracted data followed two distinct streams: (1) quantitative descriptive analysis for RQ1 and (2) a qualitative analysis informed by grounded theory coding techniques for RQ2 and RQ3.

For RQ1 (the landscape of research), we synthesized the structured data from the extraction form. This allowed us to characterize the literature and identify publication trends, dominant research areas, and key methodological parameters of the included studies. For RQ2 (Benefits and Challenges) and RQ3 (Gaps and Future Research), we employed a hybrid thematic analysis inspired by grounded theory principles [109]. Our approach combined a deductive framework aligned with our research questions and an inductive, iterative coding process to ensure both theoretical coherence and openness to emerging insights. The analysis progressed through the following four phases:

1. **Phase 1: Deductive Framework Application.** The deductive component of our analysis was embedded directly into the data extraction form. The form included separate columns corresponding to our three high-level themes: `Benefits`, `Challenges or limitations`, and `Gaps or future research`. This design ensured that as data was extracted, it was immediately structured according to the top-down deductive framework derived from our RQs.

2. **Phase 2: Open Coding.** Within each main theme, two researchers independently conducted open coding to identify and label discrete concepts emerging from the text segments. Following grounded theory conventions, this stage focused on breaking down the data into meaningful units that captured distinct ideas. Codes were continuously compared and refined in an iterative fashion – known as constant comparative analysis – to ensure conceptual clarity and consistency across studies. This phase was conducted iteratively until all the data within each thematic column were coded.
3. **Phase 3: Axial Coding.** After open coding, the researchers examined relationships and dependencies among codes to link related concepts and organize them into higher-order categories.
4. **Phase 4: Refinement and Validation.** The final step involved organizing the developed categories within each of the three main themes to ensure they were distinct, comprehensive, and accurately represented the underlying data, forming a coherent thematic structure.

Throughout all coding stages, three major consensus meetings served as critical quality assurance checkpoints, ensuring the consistent application of the coding protocol. In line with our commitment to transparency and replicability, the final codebooks and the fully annotated data spreadsheet are publicly available in our online repository[6]. The resulting coding schemes are detailed in Section 5.2. These schemes provide the structure for the synthesized findings addressing RQ2 and RQ3, which are presented in Section 6.

## 5. Results

### 5.1. Bibliometrics of qualified sources

Given the topic being highly up-to-date and dynamically advancing, the vast majority of papers that qualified for data extraction and subsequent analysis were published in 2024 or 2025. Only three papers dated back to 2023 (Figure 2). Roadmaps constituted 35.7% of processed sources, whereas secondary studies – 64.3%. On top of that, the quality of the sources could be regarded as high. All 10 roadmaps and 16 of the 18 secondary studies were published in scientific journals. Consequently, 92.9% of the sources (from both the roadmap and secondary study categories) were journal articles. The remaining two publications were published in conference proceedings. All the papers had DOIs assigned.

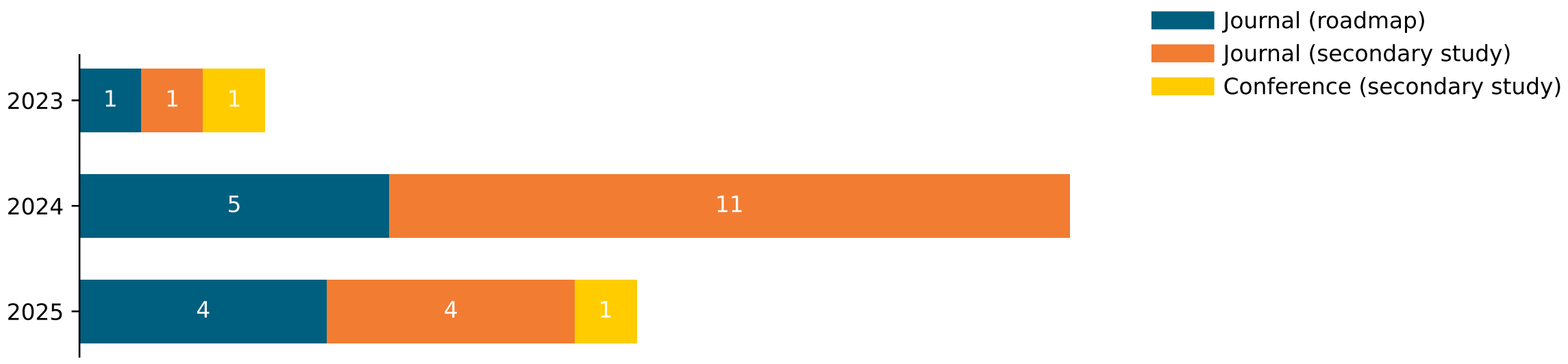

Figure 2. Papers analyzed: Publication years and types of venues

[6]https://github.com/przybylek/GenAI4IS

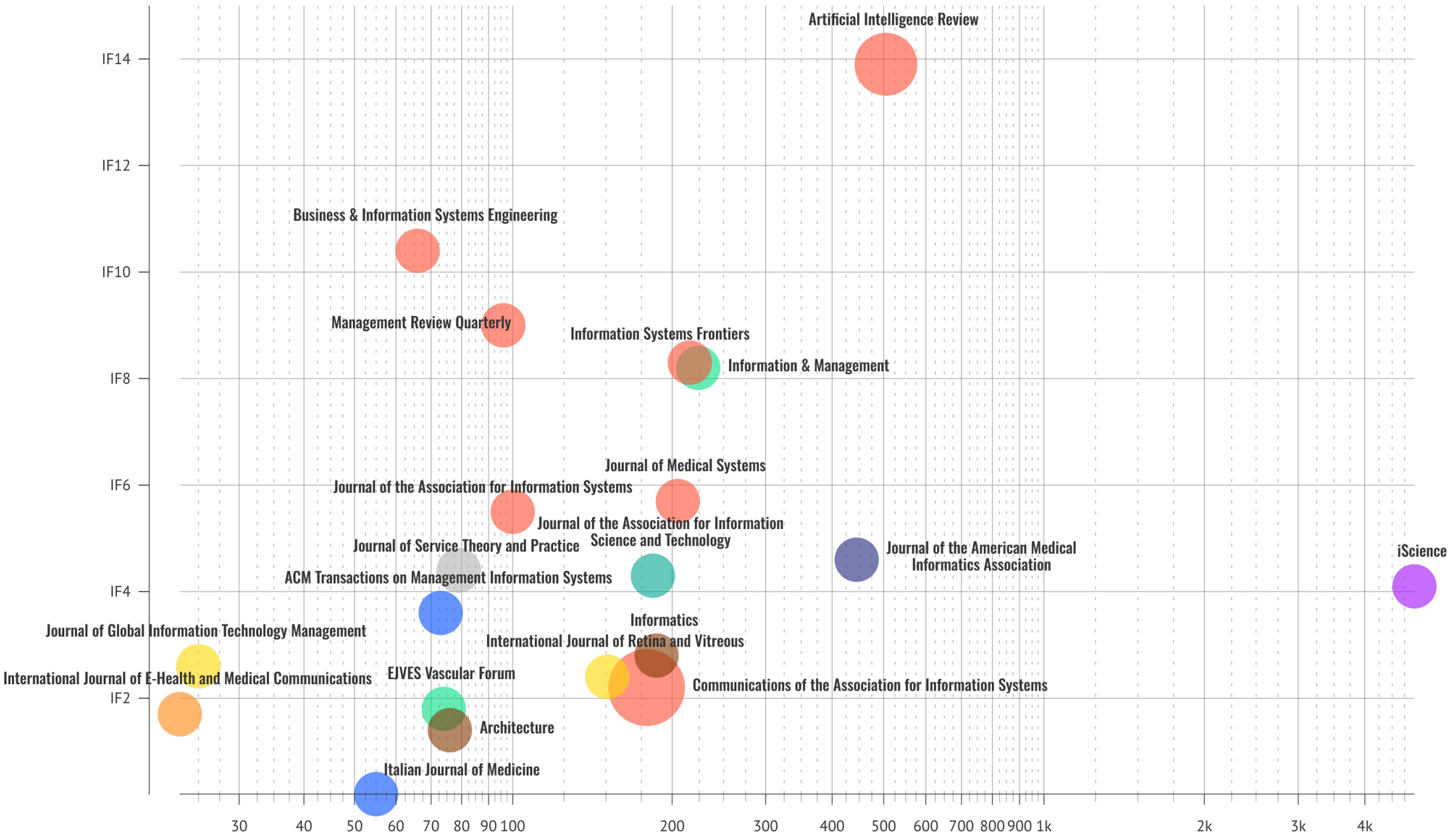

Figure 3. Journal distribution in terms of journal impact factors and citable items

Out of the journal papers, 23 came from scientific journals ranked by the Journal Citation Reports (2024 edition). In total, 42.3% of the journals in which the articles were published were classified in Q1, 19.2% in Q2, with 19.2% and 7.7% in Q3 and Q4, respectively. It is shown in Figure 3 ($X$ axis indicates the citable items; $Y$ axis the impact factor; the bubble size represents the number of sources included in our review published by a given journal). It is Springer Nature that is the entity most commonly responsible for publishing papers that were found relevant for this study (6 occurrences), with the Association for Information Systems listed in the second place (4 occurrences from JCR + one outside of JCR). In fact, the *Communications of the Association for Information Systems* was the most common journal in our analysis (3 publications), and the *Artificial Intelligence Review* followed with two published papers. The latter also had the highest Impact Factor among the group. With one exception (i.e., *iScience*), the journals were low- to medium-scale, with the number of citable items across two years ranging from 23 to 505.

The papers qualified for data extraction covered the bulk of universally recognized IS domains. As far as the application sectors are concerned, the substantial part of distinct aggregates of industries and industry groups covered by the International Standard Industrial Classification of All Economic Activities (ISIC) [110] is screened by the studies. As a matter of fact, it is the Human Health and Social Work Activities (listed within ISIC classification as sector Q) that was the most universally covered by the studies (Figure 4) – with 15 different articles analyzing or providing implications for this sector. Among other sectors that stood out, one should highlight (1) Information and Communication; (2) Professional, Scientific and Technical Activities; as well as (3) Education. The former covers, *inter alia*, a wide range of IS-centric activities related to software development, IT services, data processing, and communicating via web portals and other IS-relevant media. On the other hand, Sector M (Professional, Scientific and Technical Activities) features market research and public opinion polling division that is enables tracking Gen AI enhancement of marketing-centric information systems.

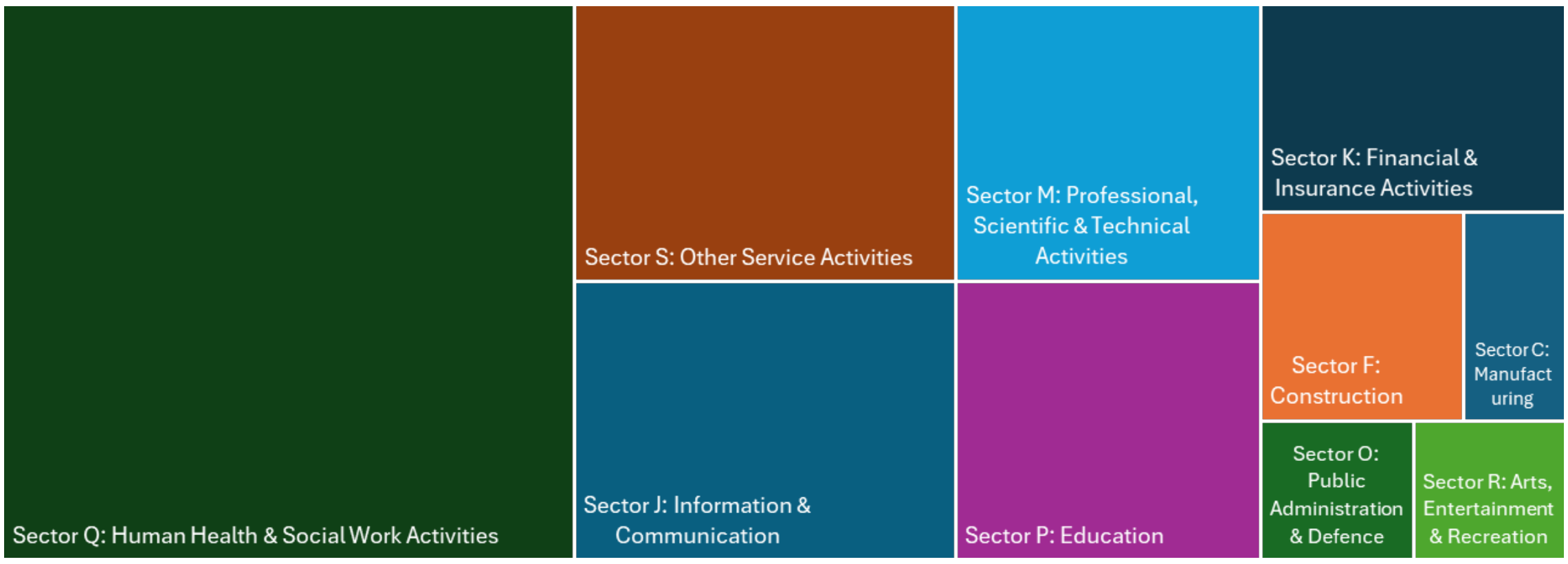

Figure 4. Papers analyzed: IS application sectors

As far as secondary studies that were covered by our research are concerned, most of them were constructed around SLR as the primary research method (Figure 5). Half of the secondary studies under analysis were confirmed to employ this method with satisfactory rigor. The scoping review followed with 5 occurrences. It should be noted that one of the studies combined SLR with Narrative Review, which is reflected in Figure 5. In 8 cases, the PRISMA protocol was followed. Most secondary studies relied on more than one database. Scopus was the predominant choice here, with Web of Science, Google Scholar, and PubMed also being used more frequently than average (see Figure 5; only databases with multiple counts are displayed). A few studies queried individual publishers' databases directly. In one case, Google Scholar was only manually searched (and so was ResearchGate) after scrutinizing reference lists and when attempting to enrich the source list with gray literature. In another case, the database-based search was intentionally narrowed down to a dozen leading IS journals. The number of primary studies qualified for subsequent analysis was reported to be between 11 and 550, with a mean of 97.9 and a median value of 41.5.

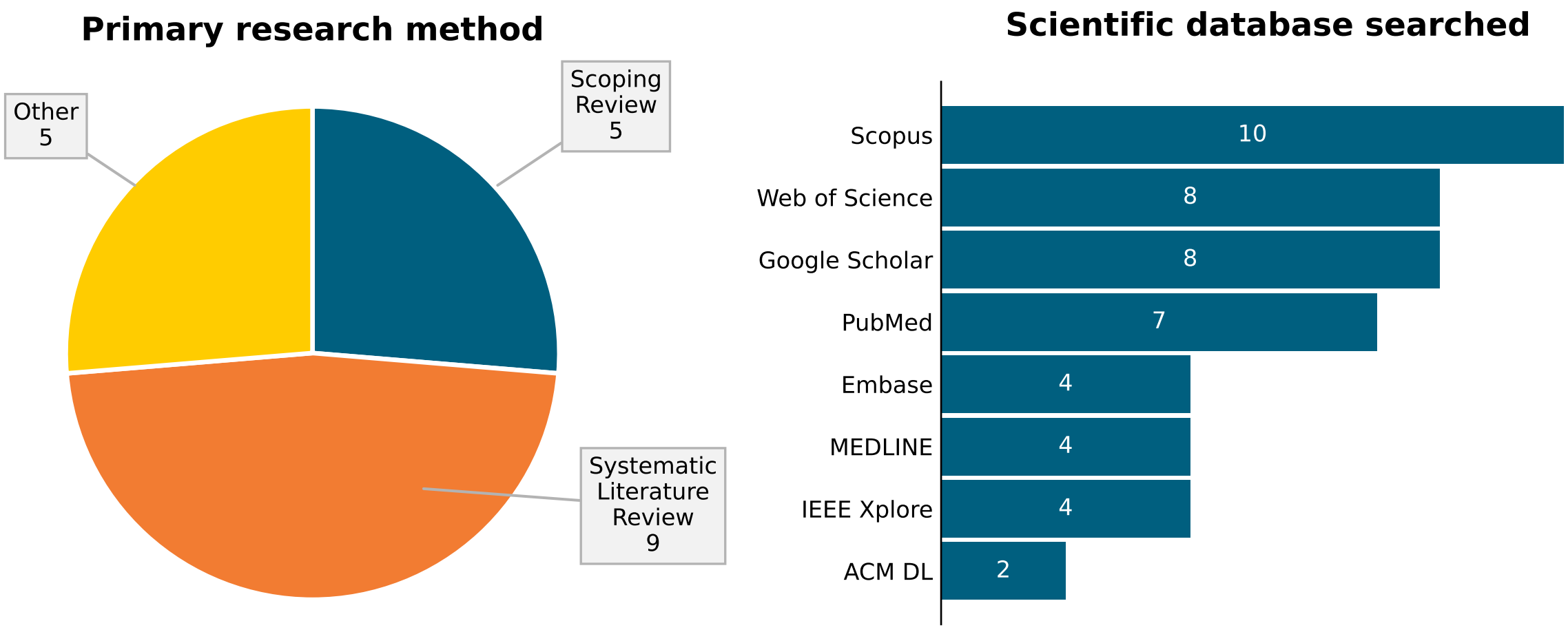

Figure 5. Method-related specificity of secondary studies

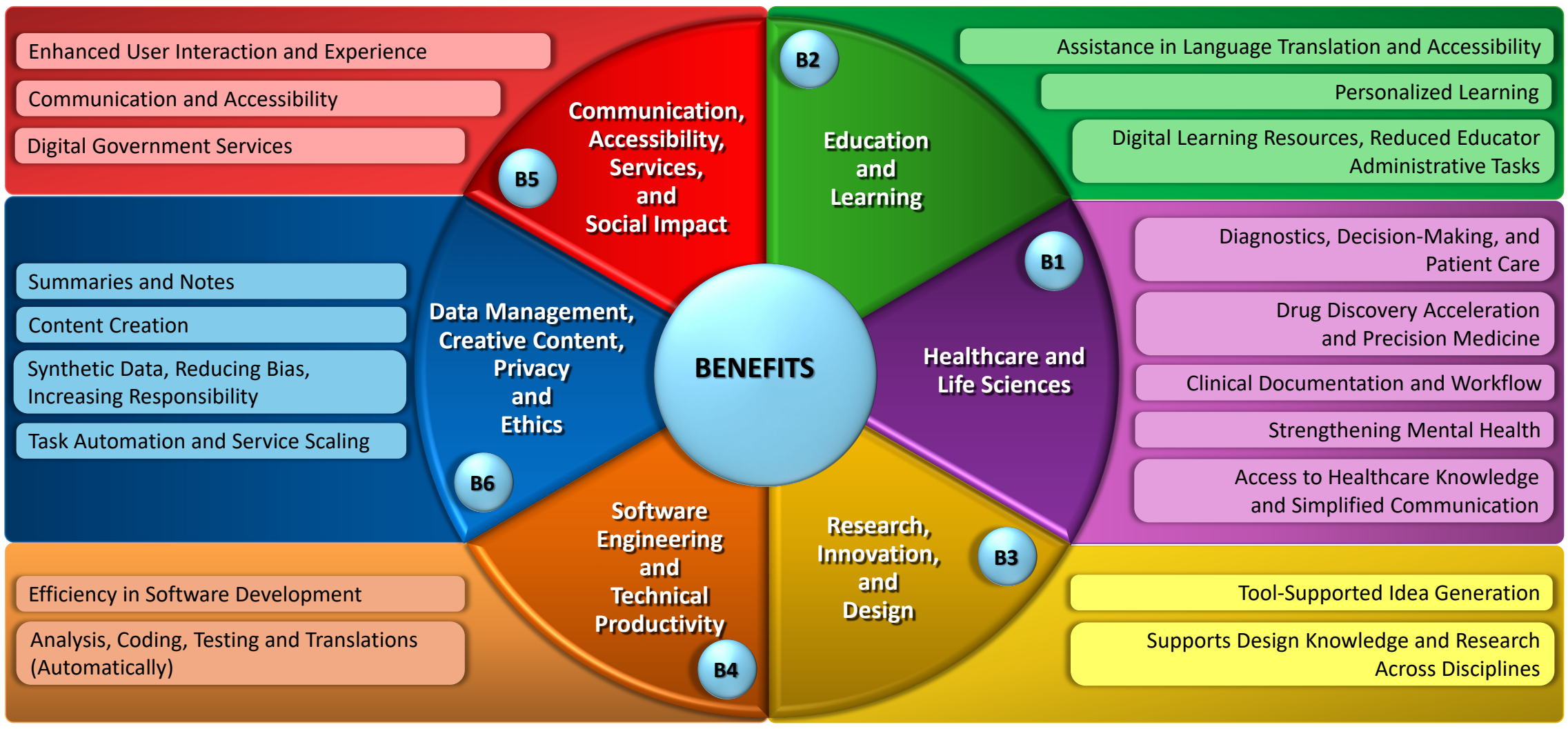

Figure 6. Theme: Benefits

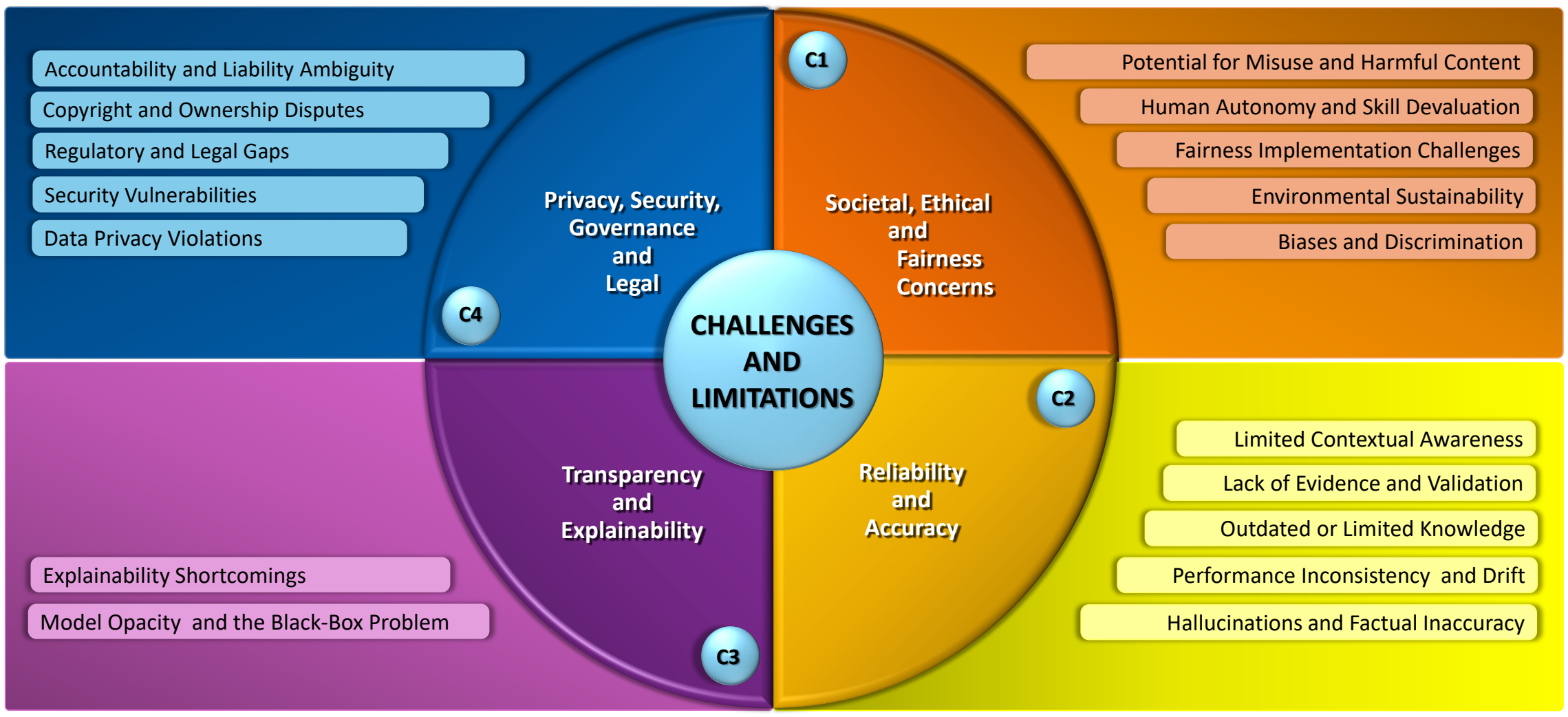

Figure 7. Theme: Challenges and limitations

### 5.2. Overview of themes and categories

Our thematic analysis of the literature identified three overarching themes that structure the current discourse on GenAI in Information Systems: (1) the transformative **Benefits** that drive the technology's adoption (Figure 6); (2) the significant **Challenges and Limitations** that temper its deployment (Figure 7); and (3) the critical **Research Gaps and Future Research Directions** that chart a path forward for the academic community (Figure 8). Each theme comprises multiple categories with associated codes that emerged inductively from our systematic analysis. Below, we present each theme along with its constituent categories. A detailed exploration of the findings is provided in Section 6.

To support the assessment of evidential strength, each theme is accompanied by a certainty-of-evidence table (Tables 9, 10, and 11) that lists every synthesized code alongside its contributing studies and their individual quality scores. These tables enable readers to directly inspect the methodological foundation of any finding of interest and to form their own judgment about the confidence warranted by the underlying evidence.

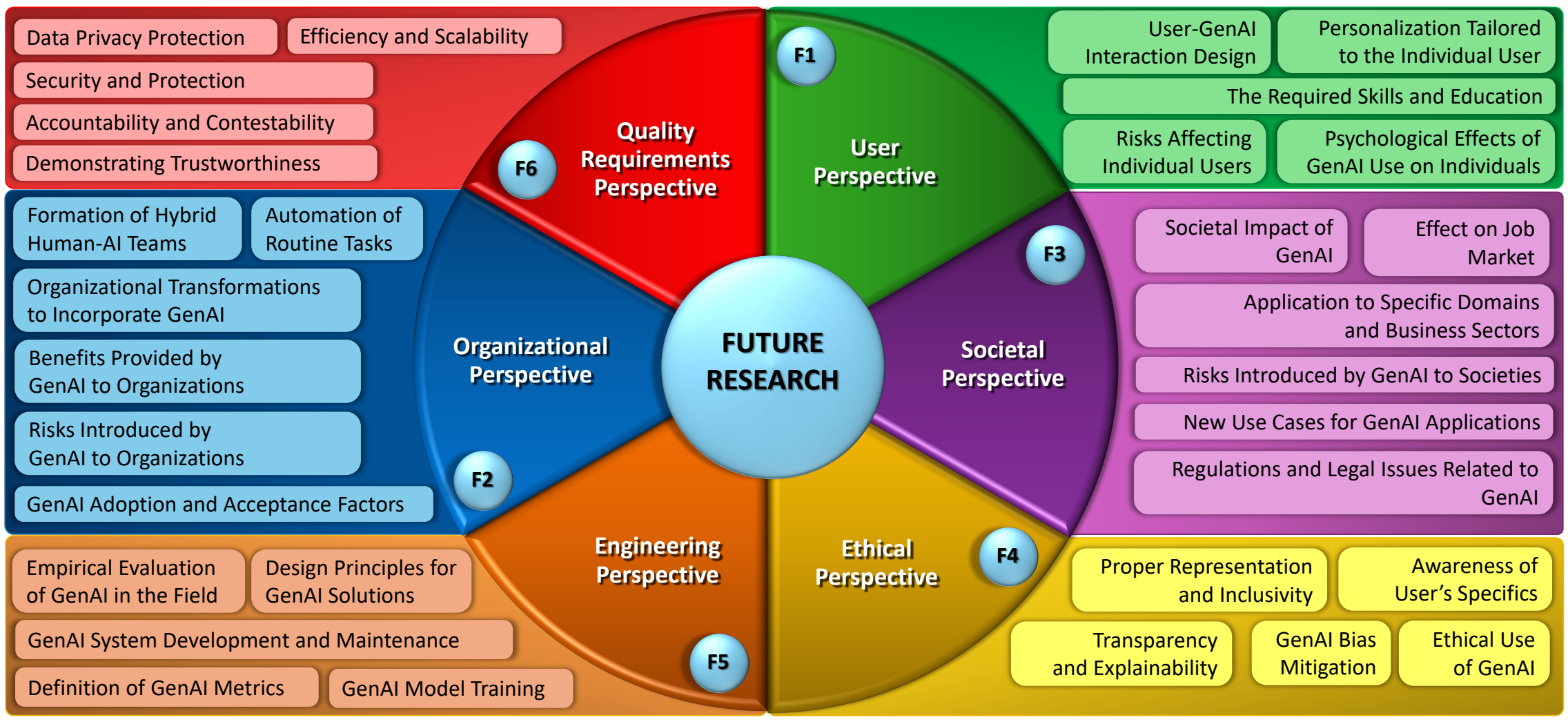

Figure 8. Theme: Research gaps and future research directions

Table 9. Certainty of evidence: Benefits

| Cat. | Code name | Contributing studies |
|---|---|---|
| B1 | Diagnostics, decision-making, and patient care | R09(1.0); S01(0.8); S02(0.8); S03(1.0); S10(0.2); S11(1.0); S16(0.0); S17(0.8) |
| | Drug discovery acceleration and precision medicine | S01(0.8); S16(0.0) |
| | Clinical documentation and workflow | S01(0.8); S02(0.8); S03(1.0); S05(0.9); S10(0.2) |
| | Strengthening mental health | R09(1.0); S01(0.8); S17(0.8) |
| | Access to healthcare knowledge and simplified communication | R09(1.0); S01(0.8); S02(0.8); S03(1.0); S10(0.2); S11(1.0); S14(0.5); S17(0.8) |
| B2 | Personalized learning | R04(0.6); S01(0.8); S10(0.2); S12(0.4); S14(0.5); S15(0.6); S18(0.4) |
| | Digital learning resources, reduced educator administrative tasks | S12(0.4); S14(0.5) |
| | Assistance in language translation and accessibility | S12(0.4); S14(0.5) |
| B3 | Supports design knowledge and research across disciplines | R01(0.7); R03(0.7); R04(0.6); R06(0.5); S02(0.8) |
| | Tool-supported idea generation | R04(0.6); R05(0.8); R06(0.5); R07(0.9); S15(0.6); S18(0.4) |
| B4 | Analysis, coding, testing and translations (automatically) | R05(0.8); S06(0.4) |
| | Efficiency in software development | R02(0.7); R05(0.8); S06(0.4) |
| B5 | Communication and accessibility | R08(0.9); R09(1.0); S09(0.6); S13(0.7); S18(0.4) |
| | Enhanced user interaction and experience | R02(0.7); R04(0.6); R08(0.9); S10(0.2); S13(0.7) |
| | Digital government services | R03(0.7); R04(0.6); R08(0.9) |
| B6 | Summaries and notes | R04(0.6); S01(0.8) |
| | Content creation | R04(0.6); R07(0.9); S04(0.5); S18(0.4) |
| | Synthetic data, reducing bias, increasing responsibility | R08(0.9); S07(0.8); S13(0.7) |
| | Task automation and service scaling | R03(0.7); R04(0.6); R08(0.9); R09(1.0); S01(0.8); S14(0.5); S18(0.4) |

Table 10. Certainty of evidence: Challenges and limitations

| Cat. | Code name | Contributing studies (quality score) |
|---|---|---|
| C1 | Biases and discrimination | R01(0.7); R02(0.7); R04(0.6); R07(0.9); R08(0.9); R09(1.0); R10(0.7); S01(0.8); S09(0.6); S11(1.0); S14(0.5); S16(0.0) |
| | Fairness implementation challenges | R01(0.7); R08(0.9); S09(0.6) |
| | Potential for misuse and harmful content | R03(0.7); R06(0.5); R07(0.9); R08(0.9); R10(0.7); S01(0.8); S09(0.6); S12(0.4); S14(0.5); S17(0.8) |
| | Human autonomy and skill devaluation | R08(0.9); R09(1.0); S01(0.8); S08(1.0); S09(0.6); S11(1.0); S12(0.4); S14(0.5); S17(0.8) |
| | Environmental sustainability | R01(0.7); R02(0.7); R04(0.6); R10(0.7); S07(0.8); S09(0.6); S10(0.2); S15(0.6) |
| C2 | Hallucinations and factual inaccuracy | R02(0.7); R03(0.7); R04(0.6); R05(0.8); R06(0.5); R07(0.9); R08(0.9); R10(0.7); S01(0.8); S02(0.8); S03(1.0); S05(0.9); S09(0.6); S10(0.2); S11(1.0); S14(0.5); S16(0.0); S17(0.8) |
| | Outdated or limited knowledge | R04(0.6); R06(0.5); S07(0.8); S10(0.2); S11(1.0); S14(0.5) |
| | Limited contextual awareness | R04(0.6); R05(0.8); S01(0.8); S07(0.8); S09(0.6); S11(1.0) |
| | Performance inconsistency and drift | R04(0.6); S05(0.9); S07(0.8); S09(0.6); S11(1.0) |
| | Lack of evidence and validation | R06(0.5); S01(0.8); S07(0.8); S08(1.0); S18(0.4) |
| C3 | Model opacity and the black-box problem | R02(0.7); R06(0.5); R08(0.9); R09(1.0); R10(0.7); S08(1.0); S09(0.6); S14(0.5); S16(0.0) |
| | Explainability shortcomings | R02(0.7); S06(0.4); S07(0.8); S09(0.6); S18(0.4) |
| C4 | Data privacy violations | R03(0.7); R08(0.9); R09(1.0); S01(0.8); S02(0.8); S08(1.0); S09(0.6); S11(1.0); S15(0.6); S16(0.0); S17(0.8) |
| | Security vulnerabilities | R03(0.7); S08(1.0); S09(0.6); S10(0.2); S16(0.0); S17(0.8) |
| | Regulatory and legal gaps | R07(0.9); R09(1.0); S08(1.0); S13(0.7); S16(0.0) |
| | Accountability and liability ambiguity | R03(0.7); S09(0.6); S17(0.8) |
| | Copyright and ownership disputes | R04(0.6); R08(0.9); R10(0.7); S09(0.6); S10(0.2); S12(0.4); S14(0.5) |

### 5.2.1. Benefits

This theme encompasses the wide-ranging benefits of GenAI across Information Systems domains. GenAI enhances healthcare and education, supports research and creative work, boosts software engineering productivity, improves accessibility and user experience, and advances data management and content generation – while underscoring the importance of ethical and responsible implementation.

**Healthcare and life sciences [B1].** In healthcare and life sciences, GenAI improves information flow, clinical decision-making, and include technologies that help patients manage and improve their health. It supports more accurate and efficient clinical workflows, accelerates drug discovery, and facilitates precision medicine through data-driven insights that enable personalized treatment. Moreover, GenAI improves documentation practices and administrative efficiency within healthcare institutions while facilitating clearer communication of medical and statistical information. Beyond clinical contexts, GenAI contributes to emotional support and mental health management, promoting well-being and improving access to healthcare knowledge across populations.

Table 11. Certainty of evidence: Research gaps and future directions

| Cat. | Code name | Contributing studies (quality score) |
|---|---|---|
| F1 | User-genai interaction design | R01(0.7); R02(0.7); R04(0.6); R05(0.8); R07(0.9); R08(0.9); R10(0.7); S01(0.8); S08(1.0); S11(1.0); S18(0.4) |
| | The required skills and education | R03(0.7); R04(0.6); R05(0.8); R07(0.9); R08(0.9); R10(0.7); S06(0.4); S12(0.4) |
| | Personalization tailored to the individual user | R01(0.7); R05(0.8); R07(0.9); R08(0.9); R10(0.7); S11(1.0); S18(0.4) |
| | Psychological effects of GenAI use on individuals | R05(0.8); R07(0.9); R08(0.9); R09(1.0); R10(0.7); S08(1.0) |
| | Risks affecting individual users | R03(0.7); R04(0.6); R07(0.9); S08(1.0); S11(1.0); S12(0.4) |
| F2 | GenAI adoption and acceptance factors | R02(0.7); R05(0.8); R07(0.9); R08(0.9); R09(1.0); R10(0.7); S01(0.8); S07(0.8); S12(0.4); S13(0.7) |
| | Formation of hybrid human-AI teams | R03(0.7); R05(0.8); R07(0.9); R10(0.7); S06(0.4); S12(0.4) |
| | Automation of routine tasks | R02(0.7); R04(0.6); R05(0.8); R08(0.9); R10(0.7); S15(0.6) |
| | Organizational transformations to incorporate GenAI | R04(0.6); R07(0.9); R08(0.9); R10(0.7) |
| | Benefits provided by GenAI to organizations | R02(0.7); R07(0.9); S01(0.8); S14(0.5); S17(0.8) |
| | Risks introduced by GenAI to organizations | R05(0.8); R08(0.9); R09(1.0); S03(1.0) |
| F3 | Societal impact of GenAI | R01(0.7); R03(0.7); R04(0.6); R05(0.8); R07(0.9); R09(1.0); R10(0.7); S01(0.8); S11(1.0); S17(0.8); S18(0.4) |
| | Application to specific domains and business sectors | R01(0.7); R02(0.7); R04(0.6); R07(0.9); R08(0.9); R09(1.0); S01(0.8); S02(0.8); S03(1.0); S05(0.9); S11(1.0); S14(0.5); S16(0.0) |
| | Regulations and legal issues related to GenAI | R01(0.7); R07(0.9); R08(0.9); R10(0.7); S11(1.0); S16(0.0); S18(0.4) |
| | Risks introduced by GenAI to societies | R02(0.7); R07(0.9); R08(0.9); R10(0.7); S01(0.8); S14(0.5); S15(0.6) |
| | New use cases for GenAI applications | R02(0.7); R04(0.6); R06(0.5); R07(0.9); S04(0.5); S12(0.4); S16(0.0) |
| | Effect on job market | R04(0.6); R07(0.9); R08(0.9) |
| F4 | Ethical use of GenAI | R01(0.7); R02(0.7); R03(0.7); R07(0.9); R08(0.9); R09(1.0); R10(0.7); S06(0.4); S09(0.6); S11(1.0); S12(0.4); S18(0.4) |
| | GenAI bias mitigation | R01(0.7); R02(0.7); R07(0.9); R08(0.9); R09(1.0); S11(1.0); S17(0.8); S18(0.4) |
| | Transparency and explainability | R01(0.7); R02(0.7); R08(0.9); R09(1.0); S01(0.8); S07(0.8); S09(0.6); S12(0.4); S16(0.0); S17(0.8); S18(0.4) |
| | Awareness of user's specifics | R01(0.7); R03(0.7); R07(0.9); S11(1.0); S17(0.8); S18(0.4) |
| | Proper representation and inclusivity | R01(0.7); R03(0.7); R09(1.0); S09(0.6); S11(1.0) |
| F5 | Definition of GenAI metrics | R01(0.7); R02(0.7); R06(0.5); R07(0.9); R08(0.9); R10(0.7); S02(0.8); S07(0.8); S13(0.7); S14(0.5); S18(0.4) |
| | Empirical evaluation of GenAI in the field | R01(0.7); R06(0.5); S01(0.8); S05(0.9); S08(1.0); S11(1.0); S17(0.8) |
| | Design principles for GenAI solutions | R04(0.6); R05(0.8); R07(0.9); S07(0.8); S08(1.0) |
| | GenAI model training | R01(0.7); R02(0.7); R03(0.7); R04(0.6); R07(0.9); R08(0.9); S02(0.8); S03(1.0); S07(0.8); S11(1.0); S16(0.0) |
| | GenAI system development and maintenance | R04(0.6); R05(0.8); S11(1.0) |
| F6 | Data privacy protection | R03(0.7); R08(0.9); R09(1.0); R10(0.7); S08(1.0); S11(1.0); S12(0.4); S17(0.8) |
| | Security and protection | R01(0.7); R03(0.7); R06(0.5); R08(0.9); R10(0.7); S11(1.0); S12(0.4); S17(0.8) |
| | Demonstrating trustworthiness | R02(0.7); R04(0.6); R08(0.9); R09(1.0); R10(0.7); S01(0.8); S02(0.8); S05(0.9); S07(0.8); S16(0.0); S18(0.4) |
| | Efficiency and scalability | R01(0.7); R03(0.7); R08(0.9); R10(0.7); S16(0.0) |
| | Accountability and contestability | R01(0.7); R07(0.9); R08(0.9); S17(0.8) |

**Education and learning [B2].** GenAI transforms education by enabling adaptive, accessible, and efficient learning. It personalizes instruction by tailoring content to individual learning styles and automates the creation of educational materials. By reducing educators' administrative and assessment workloads, GenAI allows greater focus on teaching and student engagement. Its translation and accessibility features further ensure equitable learning opportunities for diverse learners, including those with disabilities, fostering more inclusive educational environments.

**Research, innovation, and design [B3].** GenAI accelerates research and innovation by supporting knowledge synthesis, ideation, and cross-disciplinary collaboration. It automates information analysis and extraction, enabling the integration of insights from multiple domains. Within design science and innovation processes, GenAI assists in developing design requirements, principles, and prototypes, supporting design thinking methodologies. By combining human and computational creativity, it enables rapid exploration of novel ideas and solutions, driving innovation on a larger scale and at a faster pace.

**Software engineering and technical productivity [B4].** In software engineering, GenAI improves productivity and software quality by automating development and testing processes. It generates and translates code, produces test cases, and supports optimization throughout the development lifecycle, reducing manual effort and accelerating delivery. GenAI-based tools enhance workflow efficiency and decision-making, allowing teams to focus on complex design and architecture tasks while improving maintainability and reducing development costs.

**Communication, accessibility, services, and social impact [B5].** GenAI enhances communication and service delivery by enabling natural, responsive, and context-aware interactions between users and systems. Automated translation, summarization, and conversational capabilities improve accessibility and engagement across diverse audiences. In both private and public sectors, GenAI personalizes and scales service delivery, enhancing inclusivity and responsiveness. Public administrations benefit from GenAI-driven translation and accessibility features that expand reach and equity in digital government services.

**Data management, creative content, privacy, and ethics [B6].** GenAI advances data management and content creation through automated summarization, organization, and multimodal generation. It enables efficient extraction of key information from complex documents and supports the production of text, image, audio, and video content, expanding creative potential across industries. Moreover, GenAI facilitates synthetic data generation to enhance privacy, mitigate bias, and support ethical AI development by providing realistic yet anonymized datasets for training and validation. Automation of routine tasks further enhances productivity and allows organizations to focus on higher-value strategic activities.

#### 5.2.2. Challenges and limitations

This theme consolidates the significant risks and obstacles that temper the adoption of GenAI. The challenges are multifaceted, spanning from the technology's inherent technical unreliability and "black-box" nature to its profound societal and ethical implications, including the amplification of bias and the potential for misuse. These issues are further compounded by a lagging legal and governance landscape that struggles to address critical gaps in privacy, security, and accountability.

**Societal, ethical and fairness concerns [C1].** This category encompasses broad societal and ethical risks, beginning with the GenAI models' propensity to inherit and amplify societal biases from training data, leading to discriminatory outcomes and representational harm. This issue is compounded by the profound challenge of implementing fairness, which is hindered by a lack of universal standards and conflicts with organizational goals. The category also covers the significant potential for deliberate misuse, where GenAI is exploited to generate harmful content like misinformation, propaganda, and deepfakes at scale. Furthermore, it examines the erosion of human autonomy and cognitive skills, such as critical thinking, due to over-reliance, and finally, it includes the significant environmental costs associated with the high energy and computational resources required for model training and operation.

**Reliability and accuracy [C2].** This category focuses on the technical shortcomings that undermine the dependability and precision of GenAI outputs, highlighting risks in real-world applications where accuracy is critical. Core challenges include hallucinations (plausible but incorrect or fabricated information), outdated knowledge from fixed training cut-offs, and a limited contextual awareness resulting in superficial or inappropriate outputs. Reliability is further jeopardized by unpredictable performance, including inconsistent results and degradation or "drift" over time. Underpinning all of these operational risks is a pervasive lack of domain-specific benchmarks and validation frameworks, which ultimately undermines confidence in GenAI's readiness for critical applications.

**Transparency and explainability [C3].** This category deals with the challenges of making GenAI systems understandable and auditable. At its core is the "black-box" nature of transformer architectures, a fundamental opacity that hinders interpretation, auditing, and trust. This problem is exacerbated by the profound immaturity of current explainability (XAI) methods, which are insufficient for revealing the rationale behind model outputs.

**Privacy, security, governance and legal [C4].** This category addresses the constellation of vulnerabilities and regulatory hurdles surrounding GenAI deployment. It details significant risks to data privacy, where sensitive information is exposed through both model memorization and system flaws, as well as the system's susceptibility to deliberate security breaches such as jailbreaking and prompt injection. Underpinning these technical risks is a significant regulatory and legal gap, where governance frameworks lag far behind the pace of technological development. This gap, in turn, creates a critical ambiguity in accountability and liability, leaving the question of who is responsible for AI-generated harm unanswered. Finally, the category examines profound challenges to intellectual property, from unresolved disputes over the use of copyrighted material in training data to the ambiguous legal ownership of AI-generated content.

#### 5.2.3. Research gaps and future research directions

This theme encompasses the identified research gaps and the prospective future research directions identified by the scholars exploring the applications of GenAI in IS domain. The analysis of GenAI influence on individuals, organizations and societies reveals numerous topics that require further exploration. The clash between new technology and existing processes and structures raises numerous ethical questions for which there are no simple answers or solutions. The wider use of GenAI solutions requires ensuring an adequate level of quality, and to this end, it is necessary to identify appropriate quality characteristics and related quality requirements. It is also impossible to ignore the perspective of the engineers responsible for developing GenAI solutions and adapting them for use in their target context.

**User perspective [F1].** This category encompasses several aspects, from designing effective means of human-AI interaction (e.g., multimodal interfaces) and recognizing user's individual needs, through risks affecting individual users and GenAI's psychological influence on humans to the skills necessary to use GenAI for particular purposes and the ways of educating people to use GenAI more effectively and safely.

**Organizational perspective [F2].** This is the perspective of a company, enterprise or other organization using GenAI in its business processes. It includes identification of the factors important to the acceptance and adoption of GenAI by the organizations, its employees and other stakeholders. The key decision is the composition of human-AI teams including the assignments of tasks and responsibilities as well as dynamics of team's operations. This requires decisions about which tasks should be automated and which left to be performed by humans. Moreover, adopting GenAI usually requires transforming the organization and changing its processes and structures. Finally, the possible benefits and threats introduced to an organization by GenAI are a viable topic of future research.

**Societal perspective [F3].** This category focuses on the consideration in what ways GenAI can and should impact societies (professions, social groups, business sectors, nations, countries). A crucial aspect identified here is the specificity of different domains and business sectors which results in the need for investigating the specific context factors and the adjustments necessary for effective use of GenAI in a given domain/sector. The identified needs for GenAI regulations as well as the related issues of copyrights/propriety rights and legal challenges also form a promising area for research. The new, previously unknown use cases of GenAI that can bring new value as well as societal risks GenAI contributes to are directions worth investigating. Finally, the impact of GenAI on job market, i.e., replacement of jobs by GenAI solutions but also the emergence of new jobs related to GenAI usage are included in this perspective.

**Ethical perspective [F4].** The ethical issues related to GenAI are widely discussed as further research topics. Many sources indicate the need for researching the ways of ensuring that GenAI outcome is transparent with respect to algorithm, data, rationale and that its meaning is well-explained. Another widely recognized ethical concern is the possible bias and the effective ways of bias mitigation. Another issue is the requirement that the GenAI solution needs to be aware of user's specifics (e.g., national, cultural, ethnic) in order to provide the outcome suitable for such user instead of, e.g., the outcome relevant only to the countries of the highest income. Also, GenAI solutions should be trained on data that mirrors diversity in the real world and can be effectively used by users of different backgrounds and abilities.

**Engineering perspective [F5].** This is the perspective of engineers (e.g., software developers, data scientists) responsible for creating GenAI solutions. It includes the core issue of GenAI model creation and training. Also, the need for metrics that capture key properties of GenAI is clearly recognized (such metrics may differ from the ones known from systems not dealing with AI or be entirely new). Evaluation and validation of existing GenAI systems is identified as a research gap – by evaluation/validation we mean applying the GenAI system to real context (business sector, organization), using real data and observing results in the long term rather than running it on a test set and computing metrics like F1. There is a perceived lack of such evaluation, which seems necessary for GenAI's adoption in real life processes, especially in domains that require reliable evidence (e.g., medicine/healthcare). More research on development and maintenance of systems with GenAI components resulting in definitions of the corresponding processes, techniques and good practices is expected, with the emphasis on design principles for such systems.

**Quality requirements perspective [F6].** This perspective encompasses several quality properties and related categories of quality requirements relevant for GenAI solutions and not explored in detail among the previous perspectives. One of them is the trustworthiness of GenAI to its users, i.e., what factors make users willing to use GenAI and consider its responses to be reliable. Other issues include data privacy and protection of users from security threats as well as various frauds and misuses of GenAI. The well-known properties like efficiency, performance and scalability are also considered important and requiring further research. Finally, the accountability of GenAI providers and the availability of channels enabling users' complaints about inappropriate GenAI output are among topics that require more investigation.

## 6. Findings

This section provides a detailed exploration of the three core themes that structure the current scholarly discourse on GenAI in Information Systems. Building upon the framework introduced in Section 5, we synthesize evidence across the selected literature to examine how GenAI is reshaping the field – from the transformative Benefits driving adoption, through the multifaceted Challenges and Limitations tempering deployment, to the emerging Research Gaps and Future Research Directions charting the path forward. For each theme, we elaborate on its constituent categories, illustrating key concepts with evidence drawn from diverse application domains.

### 6.1. Benefits

GenAI represents a significant step forward in the development of modern digital technologies. Its applications cover a wide range of fields, from business and the creative industry to education and medicine. By leveraging the capabilities of automatic content generation and data analysis, GenAI contributes to process optimization, increased efficiency, and drives innovation in various areas of human activity.

#### 6.1.1. Healthcare and life sciences [B1]

**Diagnostics, decision-making, and patient care.** GenAI enhances diagnostic accuracy and clinical decision-making through multiple mechanisms. LLM-based models such as GPT-4 integrate extensive medical literature to recommend relevant examinations and provide evidence-based diagnostic insights [111], while also improving medical imaging for disease detection and monitoring [112]. NLP solutions automatically extract data from medical documents, identify patient conditions and disease severity, and strengthen interdepartmental communication and care coordination [113]. GenAI models further mitigate diagnostic bias through multimodal data integration and advanced machine learning techniques [114], with applications in triage and diagnosis support [115], retinal healthcare [116], dental telemedicine, and chest radiograph interpretation [117]. The technology also delivers high-quality semantic health information [118] and supports the development of personalized treatment plans through the analysis of large patient datasets [114].

**Drug discovery acceleration and precision medicine.** GenAI accelerates drug discovery by predicting drug properties and supporting genomic research [111]. It enhances the design and synthesis of novel compounds, expands and optimizes compound libraries, and enables the creation of molecules with targeted therapeutic properties [114]. By reducing the human, material, and financial resources typically required in traditional drug development, GenAI advances precision medicine through the rapid identification and optimization of candidate drugs tailored to specific patient populations and disease characteristics.

**Clinical documentation and workflow.** GenAI demonstrates enhanced efficiency in medical documentation through literature synthesis and data organization, handling diverse data types including textual patient records, diagnostic reports, research papers, medical imaging data such as MRIs and CT scans, voice recordings, and biomarkers [111]. ChatGPT offers substantial potential for automating administrative tasks [116] and supporting clinical documentation processes [115], while NLP-based technologies improve interdepartmental communication, coordinate patient care, and ensure appropriate follow-up care [113].

According to [119], research reports significant efficiency gains, with AI systems significantly reducing documentation time, particularly in ambient intelligence systems and complex clinical cases. These time savings directly impact clinician workload and patient care availability, offering a promising means to reduce healthcare professional burnout. Six of nine studies reviewed found that AI-generated documentation met or exceeded traditional standards. Healthcare professionals also reported improved usability and reduced cognitive load, supporting broader adoption of AI-assisted documentation. Additionally, NLP-based systems function as virtual assistants for health professionals, streamlining both clinical and administrative workflows [113].

**Strengthening mental health.** GenAI demonstrates strong potential to enhance mental health by providing accessible emotional and psychological support. Chatbots and conversational interfaces offer patients guidance and reassurance, delivering benefits across educational, healthcare, and broader social contexts [112]. GenAI facilitates preliminary patient consultations and psychological assistance, helping patients manage the psychological stresses associated with illness [111]. Research highlights its effectiveness in improving holistic understanding, reducing workload for mental health professionals, mitigating loneliness, and reducing the emotional burden on patients [117], collectively contributing to better mental health outcomes and quality of life through scalable, accessible support systems.

**Access to healthcare knowledge and simplified communication.** GenAI improves access to healthcare knowledge and facilitates the communication of complex medical information, enabling patients to obtain reliable insights more quickly and make informed decisions [112]. It simplifies medical terminology and statistical data, providing patients with foundational knowledge before consultations and enhancing their understanding of medical results [111]. GenAI also strengthens doctor–patient interactions through preliminary consultation tools and clearer explanations, while applications such as ChatGPT show strong potential for patient education [111, 115, 116].

NLP-based technologies act as virtual assistants, offering patients information and support with tasks such as planning, follow-up, and scheduling [113]. They provide high-quality, semantically rich health information by simplifying medical texts, conveying disease information effectively, and addressing low-risk health queries [118]. GenAI further enhances accessibility by translating medical reports into plain language, generating personalized health guidance, and supporting lifestyle interventions [117, 120]. In dental telemedicine, its multilingual capabilities improve scalability and facilitate effective consultations [117], helping democratize access to healthcare knowledge across diverse populations and literacy levels.

#### 6.1.2. Education and learning [B2]

**Personalized learning.** GenAI supports personalized learning through digital teaching assistants and the creation of supplemental materials such as teaching cases and recap questions [121]. In medical education, it enables advanced training with real-time simulations [111] and acts as a virtual assistant that generates educational content and personalized study plans [113]. ChatGPT further adapts content delivery to individual needs, fostering active engagement, self-paced learning, and deeper understanding of the subject [122].

GenAI also promotes equitable education by providing flexible, efficient and cost-effective learning opportunities. It offers instant feedback and explanations that improve self-directed learning and curiosity [122], while facilitating rapid information access and innovative teaching approaches [120]. AI-driven tools, such as chatbots, enhance these experiences by supporting creativity, automation, personalization, collaboration, multimodal content creation, and accessibility [123, 124]. Collectively, GenAI advances educational systems that adapt to individual learner profiles, preferences, and learning trajectories in diverse contexts.

**Digital learning resources, reduced educator administrative tasks.** GenAI streamlines the creation of digital learning resources while significantly reducing the administrative workload of educators. It enables the development of diverse and engaging materials that accommodate different learning styles and enhance instructional quality [122]. GenAI reduces the time spent on routine tasks by automating tasks such as generating multiple-choice questions, planning lessons, and supporting technology-based teaching [122]. It also assists in creating new educational content and exam questions, with automatic scoring and grading capabilities [120], allowing educators to focus on higher-value instructional activities and student engagement.

**Assistance in language translation and accessibility.** GenAI enhances accessibility and inclusivity in education by supporting language translation and adapting content for diverse learner populations. It streamlines classroom tasks such as lesson planning and technology-based instruction while providing instant answers that promote self-directed learning and exploration [122]. GenAI also assists in writing assignments, developing research papers, and generating educational materials and exam questions [120], ensuring that learning resources are accessible to students with different linguistic and accessibility needs, including those with disabilities.

#### 6.1.3. Research, innovation, and design [B3]

**Supports design knowledge and research across disciplines.** GenAI demonstrates transformative potential in improving productivity, decision-making, and economic value across business sectors and research disciplines [38]. It improves efficiency [125] and supports process discovery by generating process descriptions that help organizations identify and analyze different workflow stages [121]. The capacity of GenAI to model complex, non-linear business processes enables its use in implementation, simulation, and predictive monitoring, while fostering innovation through new business ideas, products, services, and models [121].

The technology reshapes organizational knowledge management by automating knowledge discovery from large volumes of unstructured, distributed data. It enhances knowledge sharing through automatic generation and dissemination of multilingual content, such as Wikis and FAQs, and delivers personalized insights to employees [121]. In design science research, GenAI supports the construction of novel IT artifacts by extracting design knowledge in the form of requirements, principles, and features, from interdisciplinary sources, making it collectively available to researchers and practitioners [121]. Integrated into design thinking and related methodologies, GenAI augments human creativity in idea generation, user needs elicitation, prototyping, evaluation, and automation [121]. Furthermore, it enables the algorithmic identification of knowledge gaps and inconsistencies, promotes new dialogic and methodological approaches, and supports the formulation of innovative research questions across disciplines [115, 126].

**Tool-supported idea generation.** According to [121], GenAI enhances idea generation across organizational functions by combining human and computational creativity. In business process management, it supports innovative process redesign and automation, driving the development of next-generation process guidance systems. It enables automated knowledge discovery, improves knowledge sharing through content generation, and maintains enterprise models at multiple abstraction levels, while supporting digital twin applications for enterprise asset management.

GenAI also delivers high-quality natural language interfaces that enhance usability and accessibility, producing optimized content for social media, emails, and reports [121]. It improves collaboration through intelligent agents, automates personalized marketing, and strengthen recommender systems through advanced personalization [121]. In design thinking and innovation contexts, GenAI supports user needs elicitation, prototyping, evaluation, and design automation [121], showcasing strong potential for human–AI collaboration that amplifies creative problem-solving [126, 127]. In architecture, engineering, and construction, it facilitates concept visualization and generation of alternative design solutions, supports data-driven decision-making and provides instant training and feedback [123]. GenAI has demonstrated exceptional creative performance, including passing university-level exams, achieved through reinforcement learning from human feedback [124, 128].

#### 6.1.4. Software engineering and technical productivity [B4]

**Analysis, coding, testing and translations (automatically).** GenAI provides enhanced support for content analysis and code generation [127], enabling developers to automate routine coding tasks and improve development efficiency. It generates and optimizes test cases to accelerate testing and meet coverage criteria [129]. Additionally, GenAI and LLMs enable seamless code translation between programming languages, reducing manual effort and supporting more efficient, interoperable software development workflows [129].

**Efficiency in software development.** GenAI improves the efficiency of software development by improving productivity, optimizing resource utilization, and reducing operational costs [8]. Its integration supports strategic decision-making and fosters human–AI collaboration that augments creative problem-solving within development teams [127]. Advanced tools such as Bard, ChatGPT, and Copilot contribute to the design of more accurate and robust software, enabling the production of higher-quality systems in shorter development cycles [129]. Furthermore, by incorporating pair programming methodologies derived from Extreme Programming, AI agents can function as collaborative team members, assisting developers throughout the software lifecycle to accelerate time-to-market and enhance code quality and maintainability [129].

#### 6.1.5. Communication, accessibility, services, and social impact [B5]

**Communication and accessibility.** GenAI facilitates the bridging of communication gaps and the delivery of tailored services across diverse audiences, promoting societal inclusion through enhanced engagement and mutual understanding. Intelligent automation enables organizations to provide personalized and adaptive services at scale, resulting in favorable outcomes for more people in society [130]. Integrating personalization and automation into complex processes enables GenAI to produce customized content and interactions that support informed decision-making and address individual needs [112, 131].

Moreover, the technology fosters social cohesion by improving cross-cultural and interdisciplinary communication, thereby enhancing societal connectivity [131]. As GenAI redefines traditional workflows and user interactions, its integration requires adaptive socio-technical frameworks that reflect evolving modes of human–AI collaboration [132]. It offers distinct advantages, including creativity, automation, personalization, multimodal content generation, and improved accessibility, while ensuring responsible adoption requires the use of explainable GenAI approaches [124].

**Enhanced user interaction and experience.** GenAI facilitates more natural, efficient, and adaptive communication between users and systems, rendering products and services increasingly intuitive and personalized. When integrated with LLMs, smart devices, and the Internet of Things, GenAI functions as an intelligent assistant that enhances individual support, productivity, and overall user experience [8]. It further enables the automated generation of personalized marketing content tailored to personality traits, such as introversion or extroversion, demonstrating superior effectiveness compared to uniform communication strategies [121].

Within service marketing and customer relationship management, GenAI supports strategic planning and operational execution by streamlining service delivery and improving customer engagement [130]. It facilitates the design of personalized service offerings and the development of targeted marketing strategies for specific customer segments, enabling scalable and adaptive service personalization through intelligent automation [130]. In parallel, NLP capabilities enable automated data extraction and analysis, while GenAI-driven conversational platforms simulate human-like interaction, contributing to widespread adoption across diverse domains [113]. As these technologies continue to reshape user interactions and organizational workflows, their integration requires flexible socio-technical frameworks that accommodate the evolving patterns of human–AI collaboration [132].

**Digital government services.** GenAI enhances digital government services through translation and accessibility features that increase service reach and improve citizen engagement. It improves efficiency across the public sector [125], supporting the digital management of non-tangible organizational assets, such as procedures, legal texts, and service documentation, throughout their lifecycles. Comparable advantages extend to the management of physical assets in Industry 4.0 environments [121].

In digital service delivery, GenAI improves the performance of existing services by producing human-like conversations with customers, providing personalized and cost-effective services [130]. It serves as a disruptive force across digital services including video streaming, recommendation agents on e-commerce platforms, online financial and banking services, education, legal, and healthcare services [130]. Governments can leverage GenAI's translation and text-to-speech technologies to broaden access to public services, while its content moderation and misinformation detection capabilities contribute to safer and more equitable digital ecosystems [130].

#### 6.1.6. Data management, creative content, privacy, and ethics [B6]

**Summaries and notes.** GenAI automates information summarization and note generation, thereby improving knowledge management and organizational efficiency. It enhances collaboration within teams by providing intelligent agents that suggest, summarize, and synthesize information based on team context, such as through automated meeting notes [121]. The technology creates summaries and notes for various applications, including medical contexts such as surgeries [111], enabling knowledge workers to extract essential insights from complex information while reallocating time to higher value analytical and decision-making tasks.

**Content creation.** GenAI accelerates content creation across multiple media formats and enables novel creative applications. It automates various tasks in marketing and media, including news writing, summarization of web content for mobile platforms, thumbnail generation, and accessibility adaptations such as text-to-speech and Braille-supported content [121]. Beyond text, GenAI facilitates multimodal content generation encompassing images, audio, and video [124], while reducing labeling requirements and expanding content creation use cases [7, 128].

However, the same capabilities also enable the production of realistic disinformation, including fake news and propaganda, which are increasingly difficult to detect. Advances in GenAI have lowered the cost of disinformation generation and introduced unprecedented personalization by adapting tone and narrative to specific audiences [121]. Moreover, GenAI can replace traditional crowdsourcing through automated annotation and execution of knowledge tasks, underscoring the need for robust ethical and governance frameworks to balance creative innovation with responsible information dissemination.

**Synthetic data, reducing bias, increasing responsibility.** GenAI facilitates the generation of synthetic data to enhance privacy, mitigate harmful biases, and promote ethical and responsible AI practices. Advances in generative models, particularly Generative Adversarial Networks (GANs), have improved the accuracy and realism of synthetic data while maintaining privacy protection [133]. In medical and scientific research, GenAI enables the creation of high-quality datasets that preserve patient confidentiality and support data-driven innovation [132].

Beyond research, synthetic data generation supports organizations in improving operational efficiency, reducing costs, and enhancing service delivery in a secure and ethical manner [130]. By enabling model training and validation without exposing sensitive information, GenAI contributes to fairer and more transparent AI systems through the creation of balanced datasets that better represent diverse populations and contexts.

**Task automation and service scaling.** GenAI automates routine tasks and enables scalable business processes, thereby enhancing productivity and allowing human workers to focus on higher-value activities. It improves organizational efficiency [125] by automating key business process management functions such as process extraction from text, event management, resource allocation, and social media operations [121]. GenAI further increases productivity by automating content creation, customer service, and code generation, with the potential to transform entire industries through large-scale process optimization [121].

In service contexts, GenAI enhances customer experience through authentic automation and cost-effective personalized interactions [130]. It enables scalable and efficient service delivery while reducing employee workload, improving both productivity and job satisfaction [130]. By automating complex organizational processes [112] and handling routine tasks such as generating budget proposals [111] and data analysis [120], GenAI facilitates greater operational agility and informed decision-making. While offering clear benefits in automation, personalization, collaboration, and accessibility, responsible adoption requires transparency and interpretability through explainable GenAI approaches [124].

### 6.2. Challenges and limitations

While generative AI offers transformative potential, its deployment introduces a complex landscape of challenges spanning technical, ethical, social, and governance dimensions. Our thematic analysis reveals four primary categories of concerns that demand careful attention to ensure responsible and effective implementation of these technologies.

#### 6.2.1. Societal, ethical and fairness concerns [C1]

**Biases and discrimination.** A predominant ethical challenge, identified consistently across the literature, is the propensity of GenAI models to perpetuate and amplify societal biases. This concern transcends domain boundaries, manifesting across general management information systems [8, 35, 38, 121, 128, 130, 131], healthcare applications [111, 112, 114, 118], and educational contexts [120]. The problem's origin lies in the models' training data, which often encapsulates historical and systemic prejudices. As Chau and Xu [8] explain,"the training data of LLMs may contain biases from various sources reflecting racial, gender, and other discriminant judgments in humans and society. Trained on these data, LLMs may inherit and amplify such biases, causing the decisions to be unfair for some social groups, communities, or societies".

The ramifications of this inherited bias are severe, leading to discriminatory outputs and representational harms that disproportionately affect marginalized groups [8, 118, 131]. This is compounded by what one study terms "exclusionary norms", where models trained on data from affluent regions neglect global diversity, thereby reflecting the "practices of the wealthiest communities and countries" and fostering cultural insensitivity [131]. The tangible risks of such biases are particularly acute in high-stakes applications. In healthcare, they can manifest as clinically inappropriate recommendations stemming from a failure to grasp linguistic or cultural nuances [111]. In organizational settings, biased algorithms can unfairly influence critical decisions like hiring and firing [112], while in education, they risk reinforcing discriminatory worldviews among learners [120].

**Fairness implementation challenges.** Addressing bias is not merely a technical problem of detection but a profound normative challenge of implementation, fraught with both conceptual and practical barriers. A primary conceptual hurdle is the absence of a universal definition of fairness, as what is considered equitable is deeply embedded in cultural, legal, and social norms [38]. This definitional ambiguity becomes particularly salient in content moderation, where, as one study highlights, "there is no universally accepted definition of what qualifies as hate speech or toxic speech" [131]. Without clear, agreed-upon criteria for what constitutes harmful content, creating globally consistent and fair moderation policies becomes exceptionally difficult.

Compounding this normative challenge are practical tensions, as the goals of fairness and equity often conflict with organizational objectives such as profitability and operational efficiency [38, 130]. These intertwined conceptual and practical obstacles mean that even when biases are identified, rectifying them in a consistent and meaningful way remains a formidable task.

**Potential for misuse and harmful content.** GenAI systems present unprecedented capabilities for generating harmful, manipulative, and malicious content at scale. The scope of potential exploitation is expansive. Laine et al. [131] provide a comprehensive catalog of deliberate misuse scenarios, ranging from the generation of "malevolent material, including spam, fraudulent reviews, or even cyberattacks on a large scale" to "creating deceptive phishing emails and malicious code". Deepfake technology represents a particularly insidious vector, enabling sophisticated identity fraud and deception, while GenAI's capacity for emotional manipulation introduces novel forms of psychological harm [35].

The propagation of misinformation and disinformation represents another pressing concern [125, 130]. GenAI models "risk blurring the line between fact and fiction, as they can rapidly disseminate false or misleading information, fake news, and malicious content, making it difficult for users to discern truth from fantasy" [131]. The consequences of such information pollution vary dramatically by domain. In healthcare contexts, misinformation can directly endanger patient safety through inaccurate medical guidance [111], while in political spheres, GenAI "can be exploited for manipulative purposes, such as the generation of propaganda or misinformation, thereby influencing public opinion and potentially harming, for example, the electoral process or other fraud and scams" [131]. Educational settings reveal different manifestations of misuse, including examination fraud and plagiarism that undermine academic integrity [120, 122].

Underlying these diverse threats is the observation that GenAI tools lack a moral compass, operating without the "intuition, plausibility, and temporal relevance" that guide human judgment [126]. This technological amorality, combined with the potential for widespread misuse, erodes societal trust in institutions and information ecosystems [131]. Consequently, it raises urgent questions about accountability and the difficulty of moderating AI-generated content at scale [111, 117, 128, 131].

**Human autonomy and skill devaluation.** A recurring theme in the literature is the concern that over-reliance on GenAI may erode essential human skills and autonomy [111, 112, 117, 134]. The primary mechanism for this is the development of a "human automation bias," where users accept AI-generated answers without critical assessment, leading to a dependency that "risks eroding skills such as creativity and critical thinking" [131]. This risk is particularly pronounced in educational settings, where the ease of content generation is seen as a threat to the development of students' independent reasoning, perseverance, and resilience [120, 122]. Beyond cognitive decline, the literature also points to the degradation of social bonds, as increased automation may reduce meaningful human interaction and lead to more profound harms, including the "loss of autonomy and dignity, dehumanization, social isolation, and addiction" [118, 130].

This erosion of autonomy is compounded by the human tendency to anthropomorphize conversational AI, which introduces distinct "psychological vulnerabilities" [131]. When users perceive AI systems as human-like, they become more susceptible to developing inappropriate dependencies and may be more likely to disclose sensitive personal information [131]. By fostering a false sense of relationship, anthropomorphism can deepen the very risks of skill erosion and social isolation previously discussed, blurring the lines between tool and companion in ways that may undermine user agency.

**Environmental sustainability.** The environmental costs of GenAI constitute an often-overlooked yet critical dimension of ethical deployment. Training and operating large-scale generative models impose substantial resource demands, resulting in significant energy consumption and carbon footprints [35,38,113,121,123,133]. As Laine et al. [131] emphasize, "the energy demands for training and operating these models contribute to resource depletion and pollution, leave a significant carbon footprint, and have high computation costs". This challenge is intrinsically linked to what Chau and Xu [8] describe as "the blessing and curse of the scaling law" – the empirical observation that model performance improves with increases in size, training data, and computational power. This dynamic creates perverse incentives that encourage ever-larger and more resource-intensive architectures, establishing a tension between performance optimization and environmental sustainability.

Compounding these concerns is the research community's limited visibility into the full scope of environmental impacts. Laine et al. [131] observe that "many environmental factors related to the operation of LLMs that are in widespread use are currently unknown", implying that documented concerns may represent merely a fraction of the true ecological cost. This opacity raises fundamental questions about the sustainability of current GenAI development trajectories and highlights an ethical imperative to systematically account for environmental impacts alongside performance metrics [123].

#### 6.2.2. Reliability and accuracy [C2]

**Hallucinations and factual inaccuracy.** The most pervasive technical challenge undermining GenAI reliability is the phenomenon of "hallucination" – the generation of outputs that appear plausible but are fundamentally incorrect [8, 35, 125, 130, 131]. This issue manifests consistently across all application domains examined, from healthcare [111, 113–119] and education [120] to general information systems [8, 121, 125–128, 130].

The consequences of such inaccuracies vary dramatically by context, with healthcare applications facing the greatest risks. Beheshti et al. [115] warn that "inaccurate or misleading information in healthcare can have severe consequences, including misdiagnoses, improper treatments, and potential harm to patients' well-being and safety". Empirical evidence substantiates this concern: Bracken et al. [119] documented "the presence of hallucinations or fictitious information in three of nine studies utilizing ChatGPT", raising fundamental questions about safe clinical implementation. The fabrication problem extends beyond medical contexts to academic and professional settings, where models have been observed generating fictitious references [120]. This phenomenon of hallucination renders human oversight and validation indispensable for any responsible application of GenAI [8, 127].

**Outdated or limited knowledge.** A structural constraint on GenAI reliability stems from the temporal constraints of their training data, which has a fixed knowledge cut-off date [120, 121, 126]. This temporal freezing creates an expanding knowledge gap as models age, rendering them increasingly obsolete for queries requiring current information. This inherent limitation is compounded by the fact that models "may not remember everything that they saw during training" [121], leading to incomplete or sparse knowledge – a particular challenge when domain-specific expertise is required [113, 118, 133].

These knowledge gaps manifest differently across domains: in healthcare as an inability to answer complex medical questions [118], in research as dependency on limited or unrepresentative datasets [133], and in general applications as an absence of temporal awareness that undermines contextual relevance [126]. Furthermore, the training process itself amplifies these limitations, as developers "often tend to rely on second-hand information from official organizations, which has a certain degree of authority but is often lagging behind" [118]. This creates a cascade of temporal delays – from the original data collection to model training to user deployment – each stage introducing additional staleness into the system.

**Limited contextual awareness.** Beyond factual accuracy, a more subtle yet critical challenge is GenAI's limited contextual awareness – an inability to interpret situational, cultural, or domain-specific nuances [111, 118, 121, 127, 131, 133]. This deficiency can render outputs technically correct yet practically useless, inappropriate, or even harmful. The risk is particularly acute in healthcare, where this limitation manifests as a failure to navigate linguistic and cultural subtleties [111] or provide genuinely personalized health information [118].

**Performance inconsistency and drift.** The reliability of GenAI is further complicated by its dynamic and often unpredictable nature. The literature highlights concerns about performance "drift", described as the unexpected deterioration of model performance over time [121], as well as inconsistent reliability across different clinical scenarios [119]. This instability also appears at a technical level, with some models exhibiting instability during training [133] or vulnerability to "semantic perturbations, whereby input with different syntax but similar meaning to the training data leads to errors" [131], and susceptibility to system crashes [118].

The combination of performance drift and inherent instability creates a particularly challenging scenario for deployment, as initial testing may not reveal latent failure modes that emerge over time or under specific conditions. This unpredictability undermines trust and necessitates constant vigilance, transforming what are marketed as autonomous systems into tools requiring continuous human oversight and validation [121].

**Lack of evidence and validation.** Underpinning all other reliability concerns is a critical meta-challenge: the lack of rigorous, evidence-based validation of GenAI systems in real-world applications. This issue is particularly pronounced in specialized domains like healthcare, where there is a "scarcity of evidence-based medical research concerning the application of LLMs in healthcare settings" [111]. The validation gap extends across multiple dimensions: absence of external validation studies, a lack of comprehensive and domain-specific evaluation metrics, and the immaturity of assessment methods and theoretical frameworks [111, 124, 133, 134].

Without established benchmarks and empirical evidence, practitioners find it difficult to delineate when these tools are productive and when they are not [126]. Compounding this challenge is a lack of reproducibility, as it may not be possible to reliably replicate tool responses through prompt engineering [126]. These validation deficiencies not only hinder safe and effective integration [111, 133] but also underscore the urgent need for rigorous, domain-specific evaluation frameworks before widespread deployment [124, 133].

### 6.2.3. Transparency and explainability [C3]

**Model opacity and the black-box problem.** The inherent opacity of GenAI models, frequently described as a "black box" problem, stems from the difficulty in interpreting how their complex transformer-based architectures arrive at specific outputs [8,35,131]. This lack of transparency is a pervasive concern that obstructs the ability to audit decision-making processes [8, 126], assess model limitations [120, 130], and ensure user control [112], eroding trust across all reviewed domains [114, 120, 130, 134]. Consequently, stakeholders are confronted with systems of immense size and "opaque behaviors" [131] whose operational logic remains inscrutable, hindering both accountability and safe adoption.

**Explainability shortcomings.** Compounding the problem of model opacity are significant deficiencies in the tools and methodologies for explaining GenAI systems [124, 129, 131, 133]. Existing Explainable AI (XAI) techniques are described as "still far from optimal," with a general consensus that available tools are marked by their "immaturity" [124]. This deficiency makes it difficult to "interpret and explain the rationale behind the decision-making process of a model" [8] or its "non-interpretable learned representations" [133]. The challenge is further amplified by the technology's scale, as "XAI for GenAI faces significant challenges due to the increasing complexity and societal reach of these models" [124]. Ultimately, without robust explainability, it is nearly impossible to debug, validate, or ethically govern GenAI systems, leaving a critical gap between their advanced capabilities and the human capacity to responsibly manage them.

#### 6.2.4. Privacy, security, governance and legal [C4]

**Data privacy violations.** A pervasive concern across all reviewed domains is the risk of unintended leakage or non-consensual exposure of sensitive, personal, or proprietary information [111, 112, 114, 115, 117, 118, 123, 130, 134]. This risk originates from multiple sources, beginning at the model level where systems demonstrate "a tendency to memorize and reproduce personally identifiable information" from their training data [131].

These inherent risks are then amplified by systemic vulnerabilities in deployment. As reported in [131], "owing to system glitches in ChatGPT, the chat logs of certain users have become accessible to others," affecting both individuals and organizations. Similarly, in enterprise contexts, organizations hesitate to use public AI tools because the prompts they submit can reveal sensitive information [125]. This risk of disclosure is further compounded by a behavioral dimension, as users are more likely to reveal private information when they anthropomorphize the technology and "treat models as if they are human" [131]. Collectively, these "privacy hazards" [131] create significant challenges for compliance with regulations like GDPR and HIPAA [114, 131].

**Security vulnerabilities.** GenAI systems are also susceptible to security vulnerabilities that expose them to malicious exploitation [113, 114, 117, 125, 134]. Laine et al. [131] document a range of adversarial attack vectors designed to compromise system integrity, including susceptibility to "jailbreaking", where prompts are used to bypass safety measures; "prompt injection" to cause malfunctions; and "data poisoning attacks" to corrupt the model's knowledge base. Such security breaches create pathways for unauthorized access, data theft, and other intentional misuses that threaten both personal and organizational security [131, 134].

**Regulatory and legal gaps,** The rapid proliferation of GenAI has created a significant governance vacuum, as existing legal frameworks are ill-equipped to manage the technology. The literature consistently notes that "current laws and regulations have become inadequate to account for new phenomena brought about by [GenAI]" [128], with the technology being adopted far faster than it can be theorized or governed [132]. This creates a profound lack of globally agreed-upon standards for ethical and safe deployment [112]. This regulatory gap poses a direct challenge for organizations attempting to ensure legal compliance [134] and for regulatory bodies tasked with developing policies to protect the public, particularly in high-stakes domains like healthcare [114].

**Accountability and liability ambiguity.** The legal and regulatory vacuum directly contributes to a critical ambiguity regarding accountability and liability. When a GenAI system causes harm through an error or biased output, there is no clear framework for assigning responsibility, leaving a notable "ambiguity over liability" [117]. This uncertainty extends to defining moral responsibility for model outputs [131] and is exemplified by the practical "warranty problem", where model suppliers refuse to provide performance guarantees, forcing adopters to shoulder the operational and legal risks [125].

**Copyright and ownership disputes.** Finally, GenAI systems raise novel challenges to established notions of intellectual property, creating unresolved disputes over both model inputs and outputs [35, 113, 120–122, 130]. On the input side, models are trained on vast datasets that may include copyrighted material without permission, potentially violating existing rights [121, 131]. On the output side, the "distinction between original and AI-generated content is blurred" [131], leading to profound "doubts about who is a legal owner of GenAI generated contents" [121]. This ambiguity fuels practical concerns about plagiarism and academic integrity [113, 122] and complicates questions of ownership and control in commercial contexts [131].

### 6.3. Future research directions

Although GenAI solutions have seen impressive development in recent years, progress in this field continues to raise new questions. This is especially true when considering a wider context beyond a technical focus. Successful application of GenAI systems and related digital transformations requires addressing many challenges and providing solutions that are much more mature than those currently available. This creates the need for future research in many areas and directions.

#### 6.3.1. User perspective [F1]

**User-GenAI interaction design.** Future research on user-GenAI interaction design in IS should prioritize the development of trustworthy, transparent, and explainable AI systems, as these themes recur across multiple sources [8, 124, 130]. In order to foster user's trust it is necessary to address issues such as hallucinations, inherited biases, and the interpretability of LLM outputs. Another identified direction is the need for personalization and adaptive design, which includes tailoring GenAI systems to individual cognitive and emotional needs, supporting diverse user routines (both professional and private), and enabling proactive emotional management [118, 127, 130]. Cultural and linguistic sensitivity is another prominent direction, with calls to localize GenAI systems and mitigate stereotypes in interactions [38, 118]. The integration of multimodal capabilities and sensory engagement further expands the design space for user experience [111, 130]. Finally, IS researchers are encouraged to explore the broader systemic implications of GenAI, including its impact on digital work, human-machine symbiosis, and the evolving boundaries of end-user computing [35, 121, 128, 134].

**The required skills and education.** The rise of GenAI is prompting a fundamental rethinking of the skills and educational frameworks required across industries. A recurring theme in the literature is the need to define and cultivate new competencies, including AI literacy and ethical usage, especially among non-technical users [122, 128, 129]. Researchers are encouraged to explore how training programs and professional development initiatives can equip both workers and educators to effectively integrate GenAI into their workflows and pedagogical practices [122, 130]. Closely linked is the transformation of job roles and labor markets, with GenAI expected to automate routine tasks, reshape existing roles, and create entirely new categories of work [121, 127, 128, 130]. Understanding which tasks are most affected and how workers can adapt is a critical area for future inquiry. Additionally, the integration of GenAI into organizational contexts – such as IT outsourcing, service marketing, and mission-critical domains like healthcare and finance – will require a reassessment of workforce capabilities and strategic positioning [35, 125]. Finally, the emergence of AI agents as team members introduces new questions about the competencies needed to work effectively alongside intelligent systems [129].

**Personalization tailored to the individual user.** Personalization in GenAI systems is an emerging topic in IS research, with multiple sources emphasizing the need to tailor interactions to individual users' preferences, values, and backgrounds [124, 127, 130]. Future studies should explore how AI-augmented services can proactively support users through personalized prompts, assistance, and emotional alignment, while also adapting to diverse cognitive and behavioral patterns. Closely related is the challenge of balancing hyper-personalization with ethical accountability, particularly in marketing and customer engagement contexts where biases may be amplified [38, 130]. Another important direction involves understanding how GenAI affects individual users – both workers and consumers – and how these technologies can enhance satisfaction, engagement, and perceived significance in digital interactions [35, 128]. Finally, personalization of explanations and communication, especially in domains like healthcare and education, is seen as a key factor for improving user comprehension and decision-making [118, 124].

**Psychological effects of GenAI use on individuals.** The psychological effects of GenAI on individuals represent a multifaceted research area within IS domain. A dominant theme across sources is the need to understand how continuous interaction with GenAI influences users' cognitive processes, emotional states, and work habits over time [35, 112, 127, 128, 130]. Researchers are encouraged to investigate both the risks of over-reliance and the potential for loss of control due to over-automation, as well as strategies to foster healthy human-AI relationships [128, 130]. Closely related is the impact of GenAI on decision-making and information-seeking behavior, which may alter users' autonomy [128]. The evolving dynamics of social and personal interactions mediated by GenAI also call for interdisciplinary, user-centered approaches that account for ethical and emotional dimensions [112, 128, 134]. Moreover, future studies should explore how GenAI systems can be designed to proactively manage emotional experiences, especially during periods of uncertainty and change [130]. Overall, this research agenda highlights the importance of capturing the nuanced and long-term psychological implications of GenAI use across diverse user populations.

**Risks affecting individual users.** The increasing integration of GenAI into everyday digital experiences raises significant risks for individual users, prompting a need for focused IS research. A recurring concern is user's privacy and data security, especially in real-world and real-time chatbot interactions where user data may be exposed without sufficient safeguards [118, 134]. Researchers are urged to develop ethical and regulatory frameworks to address these vulnerabilities. Additionally, the amplification of cyber threats through GenAI and the spread of AI-generated disinformation call for advanced countermeasures and real-time detection systems [121, 125]. Another identified risk concerns over-reliance on GenAI and the psychological consequences of excessive automation, including loss of control and technological resistance [122, 128]. Understanding these risks is essential to ensure safe and responsible use of GenAI technologies.

#### 6.3.2. Organizational perspective [F2]

**GenAI adoption and acceptance factors.** A prominent direction for future research on GenAI adoption and acceptance in IS involves understanding the trust-related factors that influence user interaction with LLMs. Multiple sources emphasize the importance of explainability, interpretability, and transparency as key enablers of trust and acceptance, particularly in domains such as healthcare and education [8, 111, 112, 133]. These factors are closely tied to concerns about bias, hallucinations, and data quality, which design science researchers are encouraged to address through novel system architectures and validation methods [8, 133]. Another widely discussed area is the organizational and cultural context of GenAI adoption, including how organizational norms, structures, and processes shape individual and collective attitudes toward automation and AI integration [35, 127, 128]. Additionally, scholars call for investigations into sector-specific adoption dynamics, such as in education [122, 128], customer service [130], and clinical decision-making [112], highlighting the need for context-aware frameworks. Finally, there is a recognized need to refine or develop new theoretical models that capture the non-linear and dynamic nature of GenAI adoption, moving beyond traditional technology acceptance paradigms [132].

**Formation of hybrid human-AI teams.** An emerging and widely discussed research avenue in the IS field concerns the formation and functioning of hybrid human-AI teams, where humans and GenAI systems collaborate in shared tasks. Several sources highlight the need to explore collaborative dynamics, including how responsibilities, decision-making authority, and handover points are distributed between human and AI agents [127–129]. This includes investigating symbiotic relationships that augment human intelligence rather than replace it, and understanding how to design collaboration frameworks that prevent over-reliance on AI while leveraging its strengths [122, 128]. The organizational implications of such hybrid teams are also significant, as GenAI adoption is expected to reshape business processes, IT capabilities, and workforce structures [35, 125]. Moreover, researchers are encouraged to examine how cultural, regulatory, and sector-specific contexts influence the integration of AI into teams [125]. Finally, the evolving composition of teams – including the competencies required to work effectively with AI agents – presents a rich area for inquiry, calling for new models of team design and skill development [129].

**Automation of routine tasks.** A key area of interest in GenAI research within IS is the automation of routine and repetitive tasks, which has implications across organizational and managerial levels. Several sources emphasize the potential of GenAI to support or fully automate tasks in domains such as education, enterprise management, and service operations, often through the use of engineered prompts and workflows tailored to specific functions [8, 121, 130]. This shift invites deeper investigation into how automation can augment human capabilities, redefine job roles, or even create new forms of work, rather than merely replacing existing ones [127]. Strategic concerns such as risk containment, accountability, and transparency in automated processes are also critical, especially in customer-facing and regulated environments [127, 130]. Furthermore, researchers are called to explore the implications of automation, including the extent of job replacement [35]. Finally, the relevance of automation extends to broader contexts such as smart cities and sustainable infrastructure, where GenAI can play a role in optimizing design, construction, and operations [123].

**Organizational transformations to incorporate GenAI.** The integration of GenAI into organizational contexts is expected to drive significant transformations in business processes, structures, and strategic models. Multiple sources emphasize the role of GenAI in revealing opportunities for process innovation and supporting process (re-)design initiatives, particularly through automation and augmentation of decision-making and resource management [121, 128, 130]. Researchers can also explore how GenAI can facilitate digital transformation across industries, including shifts from low- to high-value services and the emergence of new business models [128, 130]. Moreover, GenAI's impact on organizational capabilities, including IT infrastructure, human resource management, and knowledge systems, calls for a rethinking of traditional organizational boundaries and roles [35, 128]. Scholars are also urged to adopt a sociotechnical perspective, examining GenAI not merely as a technical tool but exploring its boundary conditions that define its presence and impact within the context of an organization [35].

**Benefits provided by GenAI to organizations.** Future research in IS should examine GenAI's impact across diverse sectors such as medicine, education, tourism and e-commerce [111, 117, 120, 128] assessing its potential benefits, e.g., the ability to personalize and augment services [117], as well as enhanced productivity, creativity, and service quality [8]. Importantly, these benefits must be evaluated alongside ethical considerations and societal implications, ensuring responsible deployment that aligns with values such as social justice and sustainability [8, 117].

**Risks introduced by GenAI to organizations.** Understanding the risks and challenges associated with GenAI adoption is essential for responsible organizational integration. Key concerns include the strategic containment of automation risks, particularly in service industries where GenAI may disrupt established marketing and operational practices [127, 130]. In high-stakes domains like healthcare, researchers are urged to address issues of accuracy, guideline compliance, and implementation challenges, ensuring safe and effective use of AI-driven tools [112, 116]. These risks highlight the need for robust governance frameworks and continuous evaluation of GenAI's organizational impact.

#### 6.3.3. Societal perspective [F3]

**Societal impact of GenAI.** The societal impact of GenAI is a multifaceted area of inquiry in IS, with researchers increasingly called to examine its implications for equity, labor markets, global development, and ethical governance. A recurring theme across sources is the need to understand how GenAI may displace or transform jobs and crowdsourced initiatives, and what welfare consequences this may entail [35, 121, 127]. These changes raise additional concerns about "societal stres" caused by job replacement [35]. Another major research direction involves the global dimension of GenAI's impact, including its role in widening or bridging the digital divide between countries at different stages of technological development, and its influence on resource allocation across the Global North–South divide [128]. The integration of GenAI into global IT management and outsourcing structures further highlights the importance of addressing regulatory diversity, cultural sensitivity, and language-specific capabilities to ensure inclusive and deployment [125]. Ethical tensions also emerge in scenarios where bias mitigation efforts may reduce profitability or create competitive disadvantages, prompting calls for frameworks that balance efficiency, fairness, and social responsibility [38, 124]. In sectors such as healthcare and education, GenAI's societal role is particularly pronounced. Researchers emphasize the need for empirical, interdisciplinary, and user-centered studies to validate its pact on public health systems and education systems [111, 117, 118]. These studies should account for cultural, linguistic, and socio-economic diversity, ensuring that GenAI technologies are designed and evaluated in ways that reflect real-world complexity. Finally, scholars are encouraged to adopt a sociotechnical systems perspective, recognizing GenAI as a deeply embedded actor within societal ecosystems, whose boundaries, interactions, and ethical implications must be critically examined [35, 112].

**Application to specific domains and business sectors.** Overall, future IS research should focus on developing tailored GenAI solutions that are not only technically sound but also ethically responsible and contextually relevant. A recurring issue related to GenAI contextual adaptation and domain-specific performance is the need to fine-tune generative models to meet the unique requirements of sectors such as healthcare, finance, education, marketing, e-commerce, tourism, and entertainment [8, 38, 121, 128, 130]. Researchers are encouraged to explore how GenAI can support enterprise management, as well as process design and re-design, contributing to operational efficiency and strategic innovation [121]. In high-stakes domains, such as healthcare and finance, the development of ethical guidelines and regulatory frameworks is essential to balance the competing demands [38, 112]. Healthcare, in particular, emerges as a focal point for future research, with calls for interdisciplinary collaboration, clinical validation, and standardized evaluation tools to assess GenAI's utility in diagnostics, documentation, and patient communication [111, 114–116, 118, 119]. In education, GenAI's potential to enhance learning experiences and develop student skills invites further investigation into pedagogical models and responsible use practices [120, 130]. Additionally, it is worth to examine how GenAI can transform public services, including government-led digital initiatives, while minimizing misuse in sensitive domains such as legal and healthcare services [130].

**Regulations and legal issues related to GenAI.** Legal and regulatory issues surrounding GenAI are becoming increasingly central to IS research, as organizations and governments grapple with the challenges of ethical deployment, data governance, and intellectual property protection. A key direction involves the development of ethical guidelines and legal frameworks tailored to sectors such as healthcare and finance, where the tension between rapid deployment, accuracy, and inclusivity creates unique regulatory demands [38]. Another prominent area of inquiry concerns the governance of bias and fairness, including how AI laws and data filtering techniques can be designed to reduce algorithmic discrimination while maintaining transparency and accountability [124, 128]. The rise of GenAI also raises complex questions about intellectual property rights, prompting calls for new metrics and legal interpretations that reflect the generative nature of these technologies [128, 130]. In the context of service industries and marketing, governments and organizations must address emerging privacy, security, and user data protection challenges, while establishing clear standards for responsible use [118, 130]. The applications of GenAI in areas such as healthcare and legal cases further underscore the need for comprehensive regulatory oversight, which requires building interdisciplinary collaborations among technologists, professionals, ethicists, and policymakers to ensure that GenAI tools are aligned with ethical principles and legal requirements [35, 114, 118].

**Risks introduced by GenAI to societies.** GenAI introduces a range of societal risks that warrant close attention from IS researchers. Key concerns include its potential to disrupt industries such as tourism, e-commerce, and healthcare, and to alter human interactions and decision-making in ways that may affect creativity, productivity, and social justice [8, 128, 130]. In sensitive domains like medicine and education, scholars emphasize the need for rigorous evaluation and standardized tools to assess GenAI's safety and effectiveness [111, 120]. Broader issues such as privacy, security [35], and the impact of automation on urban environments [123] also require investigation to ensure GenAI supports sustainable and equitable societal development.

**New use cases for GenAI applications.** Exploring new use cases for GenAI applications is a dynamic and promising research direction in IS. Scholars are increasingly interested in how GenAI can enhance existing research methods or even propose novel ones, potentially transforming the way knowledge is produced and disseminated [8, 126]. In design science, GenAI is seen as a tool to foster creativity in the development of new IT artifacts [121]. This opens the door to new genres of academic publication and alternative theorizing processes, which may challenge traditional norms and encourage methodological diversity [126]. Beyond academia, GenAI is being applied to address global grand challenges, such as environmental protection and the Sustainable Development Goals, by expanding modes of explicit knowledge production and improving efficiency in problem-solving [128]. In management and cybersecurity research, generative algorithms are used to simulate cyber-attacks, generate marketing content, and explore social media dynamics, demonstrating their versatility in both analytical and creative tasks [7]. In education, GenAI supports digital pedagogical innovations, blending AI-driven assistance with traditional teaching to create modern, adaptive learning environments [122]. In healthcare, emerging use cases include remote patient monitoring and predictive analytics, which promise to enhance care delivery and operational efficiency [114]. All such new and innovative applications require further research to enable both efficient and safe use of GenAI.

**Effect on job market.** GenAI is expected to significantly reshape the job market, particularly by automating routine tasks and transforming roles in information-intensive sectors [121, 128]. While some jobs may be replaced, new roles will likely emerge that focus on collaborating with AI systems and leveraging human-specific skills [128]. In consumer-facing services, GenAI may alter employee responsibilities and require reskilling to adapt to changing workplace dynamics [130]. As the impact of GenAI on the job market can result in tensions and resistance in societies, this topic is of particular interest to researchers.

#### 6.3.4. Ethical perspective [F4]

**Ethical use of GenAI.** Future research on the ethical use of GenAI should aim to balance innovation with societal responsibility. Achieving this will require interdisciplinary research focusing on both systemic frameworks and user-centered practices. Key priorities include developing comprehensive ethical and regulatory guidelines to address privacy, data security, and fairness, alongside investigating user training to mitigate misuse, especially by non-technical audiences [38, 118, 122, 128, 129]. Researchers should also empirically evaluate social, psychological, and economic impacts while exploring complementary ethical approaches, taking into account diverse stakeholder perspectives [8, 112, 131]. The emergence of explainable GenAI (GenXAI) highlights a need for transparent, contextualized explanations and effective bias mitigation in high-stakes domains [124]. Across sectors – from IT outsourcing to education – future work must ensure ethical integration of GenAI technologies, promoting social justice and operational accountability as commercial adoption accelerates [35, 125, 130].

**GenAI bias mitigation.** Future research on mitigating bias in GenAI, especially LLMs, calls for multidisciplinary approaches focusing on dynamic, scalable, and context-aware methods. These should exploit various approaches, from real-time user feedback systems to resource-efficient longitudinal studies for bias detection and mitigation [38]. Understanding bias requires addressing how foundational model biases propagate into downstream applications where speed and scalability often challenge fairness objectives [38, 112]. Dealing with the problem of bias also requires prior research on bias sources and types [8,38,128,130]. Such research could be supported with explainable AI [124]. Understanding bias across cultural and national contexts requires balancing localized adaptations with global standards [38]. Ethical governance is critical for effective bias mitigation, requiring multi-stakeholder frameworks emphasizing transparency, accountability, explainability, human oversight, privacy, and the right to contest AI outcomes, all embedded in informed regulatory standards and laws tailored for GenAI [117, 118, 128].

**Transparency and explainability.** Personalization of explanations based on user expertise and context is a critical direction for enhancing accessibility and ensuring that diverse audiences could meaningfully interpret model outputs [124, 130, 131]. Design science perspectives can support the development of innovative techniques that strengthen user understanding, adoption, and trust in these systems [8, 117]. In order to improve interpretability while maintaining performance, explainable AI methods should be considered [8, 38]. Apart from technical solutions, developing comprehensive ethical frameworks and guidelines addressing transparency should be considered [122]. The explainability is highlighted as a crucial factor for GenAI's healthcare applications, where future research studies should examine how explainability affects clinician reliance on automated diagnostics [111, 112, 114, 133].

**Awareness of user's specifics.** Research on user-specific awareness in GenAI should explore applications where GenAIs interact with diverse users in various contexts, such as employee recruitment, credit scoring, customer service, sentiment analysis, and content recommendation [38]. The lack of such awareness can affect fairness of such GenAI solutions. Integrating generative models within outsourcing and localization strategies demands sensitivity to local regulations, cultures, and language capabilities [125]. Localization extends to the adaptation of GenAI tools for new languages and new communication contexts, putting stress on appropriate translation and cultural relevance [118]. This is particularly relevant in healthcare applications that require careful evaluation across linguistic, cultural, and socio-economic landscapes, with attention to environmental sustainability [117]. The design of GenAI must consider whether models are built for static or dynamic contexts, further complicating the awareness issue and inducing likely fairness and transparency challenges [128]. Finally, future work should develop explanatory frameworks attuned to user needs and ethical-social factors, as generative models mature [124].

**Proper representation and inclusivity.** Inclusivity-oriented research is needed to mitigate data biases that result in unequal service or treatment quality and to strengthen data governance practices [38, 112, 118]. Similarly, the perspectives of underrepresented users on AI ethics – especially regarding explainability and perceived fairness – require systematic investigation [131]. Scholars should examine how GenAI can enable culturally adaptive and inclusive digital localization processes that serve global audiences while preserving local authenticity [125]. Future research on proper representation and inclusivity in AI faces a key challenge in developing ethical and fairness-aware frameworks for LLMs across domains such as healthcare, finance, and marketing, where rapid deployment often conflicts with equity objectives [38]. Research should further explore how to distinguish between purposeful and unintended differentiation in such systems to prevent inequitable outcomes for marginalized groups [38].

#### 6.3.5. Engineering perspective [F5]

**Definition of GenAI metrics.** A strong message is voiced about the need for new metrics that would enable evaluation of GenAIs. Such dedicated metrics could provide a base for specialized tools to assess benefits and risks of GenAI models in various fields [120]. Traditional measures, such as accuracy, precision, and recall are insufficient for GenAI tasks. Therefore, new metrics incorporating helpfulness, harmlessness, honesty (HHH), security, and standardized validation on independent datasets are necessary [8, 115, 133]. These should address the impact of GenAI on individuals, workers, and organizations, as well as the inter-organizational impacts [35]. Regarding individuals, measuring GenAI's influence on cognitive aspects such as questioning, rigor, and clarity is underexplored and warrants specific evaluative frameworks [126]. Accuracy assessment, especially concerning AI hallucinations in generative models, remains a challenge and calls for dedicated reliability metrics [130]. New metrics are also needed to measure fairness and diversity in LLM-driven systems [38] and explainable AI could help in their application [124]. Also intellectual property rights protection demands comprehensive metrics to safeguard legal and ethical standards in GenAI applications [128]. Lastly, refinement of theoretical frameworks on technology adoption is needed to better capture the nuances of GenAI integration and acceptance in diverse environments [132].

**Empirical evaluation of GenAI in the field.** The empirical evaluation of GenAI systems calls for developing resource-efficient longitudinal designs that could monitor how GenAI tools mitigate bias over time while accounting for expertise gaps and computational constraints [38]. It should definitely include user-centered aspects of GenAI systems, particularly the ethical and privacy implications of conversational models. Future studies should move beyond simulated testing to investigate real-world user interactions, collecting authentic behavioral data while safeguarding user privacy and ensuring ecological validity [134]. There is also a recognized need to empirically assess whether GenAI-based tools genuinely enhance academic rigor, questioning, and clarity in scholarly inquiry [126]. In the healthcare domain, clinical trials, observational studies, and cross-institutional collaborations could help determine the real clinical utility of language models [111]. Achieving this goal requires engaging healthcare professionals directly involved in documentation and decision-making processes, as well as expanding the scope of research to encompass diverse medical contexts [119]. A coherent approach to validating AI systems in health communication must also consider technical accuracy, patient satisfaction, and public health outcomes through real-world trials [118]. GenAI-driven personalized healthcare services should be studied across cultural and socio-economic boundaries, evaluating their sustainability and impact on equitable care delivery [117].

**Design principles for GenAI solutions.** Research on design principles for GenAI should first address the question of effective design principles that guide GenAI development holistically, integrating human-centered and technical perspectives [121]. An important research direction explores how to design and test GenAI-based systems with varying levels of automation to optimize human benefit considering individual diversity [127]. This personalization-oriented approach complements efforts focused on defining the most effective design processes for creating GenAI that operates collaboratively with humans, balancing autonomy and human oversight [128]. Another theme is the simplification of model architectures, improving data quality, and implementing standardized validation procedures in a strive to ensure that GenAI systems are reliable, transparent, and suitable for critical settings such as healthcare [133]. Moreover, privacy-oriented design emerges as a necessary complement to usability and transparency research. Addressing gaps in privacy-aware design demands systematic exploration of how user-privacy principles can be embedded throughout the development lifecycle, from early prototyping to deployment. This includes not only identifying potential risks but also developing practical techniques, frameworks, and design guidelines for privacy-sensitive interfaces like chatbots [134].

**GenAI model training.** Future research on GenAI model training converges around several key domains: model adaptation, bias mitigation, domain specialization, and ethical application. One major direction involves developing strategies for effective model fine-tuning to balance accuracy and generalization, without overfitting or sacrificing scalability [38]. Closely related to these concerns is the effort to fine-tune or adapt models for specific domains and contexts [8, 121]. Research should examine whether GenAI is best suited for static or dynamic environments, and how models can evolve continuously without structural conflicts during extended use [128]. The integration of GenAI in outsourcing sector necessitates to train the models considering global regulatory variations, cultural differences, and language specific capabilities while addressing efficiency and innovation [125]. The challenge of preventing bias in algorithms and data used to train models persists as a critical focus area [38, 130]. Healthcare is an exemplary target area for which domain-specific improvements in model training should be considered. These include enhancing model accuracy, reliability, and robustness for safe clinical deployment [114], guidance by up-to-date medical standards [116], standardized validation and diverse training datasets representing multiple disease categories for supporting reproducibility and generalization of research [133], as well as advanced multimodal learning integrating textual and visual medical data to improve diagnostic understanding [118]. As an alternative (or a support) to domain-specialized models, Retrieval-Augmented Generation could be considered [115]. As regards ethical application, guidelines for the responsible integration of LLMs should be developed for high-stakes sectors such as healthcare and finance, emphasizing the balance between rapid deployment, accuracy, and inclusivity with regulatory considerations [38].

**GenAI system development and maintenance.** A promising future research direction is the exploration of GenAI's role in design science research to enhance creativity in developing new IT artifacts which could lead to advancing the theoretical and practical foundations of IT development [121]. A related focus is the development and testing of partially automated tools aimed at maximizing human benefit. Specifically, advancing the technical integration of AI within robotic process automation (RPA) tools could lead to more sophisticated, adaptive automation solutions in organizational contexts [127]. In terms of system integration, future work should target addressing interoperability, usability, and compliance challenges. For instance, in the case of existing healthcare information systems, including electronic health records (EHRs), it would enable GenAI to augment clinical workflows and decision-making [118].

### 6.3.6. Quality requirements perspective [F6]

**Data privacy protection.** Data privacy protection should be considered of primary importance in future research on GenAI due to the risks and harms, especially the ones arising from real-time interactions with chatbots. Current studies often rely on simulated data, leaving a gap for research based on actual user data in real-world settings [134]. Future work should focus on developing user-privacy-centric chatbot designs, incorporating robust privacy safeguards and intellectual property protections [130, 134]. Personalization and user interaction improvements, particularly in health communication, should ensure that privacy and consent safeguards are in place to deem the related GenAI solutions trustworthy. There is a need to study GenAI telehealth applications across diverse cultural and socio-economic contexts to evaluate informed clinical decision-making and sustainability [117, 118]. The broader societal impact includes transformative changes in workforce and operational processes across sectors. This calls for renewed research into end-user computing and examination of societal issues like privacy and security in the pervasive use of GenAI [35, 125]. The growing sophistication of GenAI-powered privacy-related cyber threats requires IT outsourcing firms leveraging GenAI to develop advanced real-time detection, response, and mitigation systems to protect sensitive data and maintain trust with clients [125]. There is a growing demand for comprehensive guidelines addressing data privacy protection that would define accountability structures involving all stakeholders to ensure safe and transparent deployment [117, 118, 122]. This could be undertaken as a part of a wider effort on data governance concerning AI, especially in healthcare, which remains an understudied area in IS research [112].

**Security and protection.** Security challenges are amplified by GenAI's capabilities, leading to sophisticated cyber threats. The development of advanced AI-driven real-time detection, response, and mitigation methods is critical, alongside protocols against attacks such as prompt injections and data poisoning. Human oversight and clear accountability frameworks are essential to secure GenAI systems effectively [117, 125]. This context necessitates a deeper understanding of broader societal impacts, especially concerning the security of business applications [35]. Furthermore, strengthening defenses against GenAI-fueled security threats is of paramount importance. As adversarial actors exploit generative tools to scale cyber attacks, new AI-driven mechanisms for real-time threat detection, adaptive response, and mitigation must be developed [125]. This effort should align with a sociotechnical assurance framework emphasizing human oversight, resilience against data poisoning and prompt injection, and clear accountability chains [117].

A closely related yet distict research direction should concern combating the misuse of GenAI. Governments must proactively prepare for the potential misuse of GenAI by establishing comprehensive regulations and guidelines that address its deployment in sensitive services such as healthcare and legal sectors. This involves creating frameworks to minimize generative tools' misuse, ensuring robust oversight and accountability. Additionally, preventative measures should be implemented to guard against malicious uses of AI systems, fostering a responsible environment that prioritizes ethical considerations and protects individuals and institutions from harm [130]. Security also gives reasons to design and enforce comprehensive frameworks to guide the responsible development and deployment of GenAI systems. These frameworks should address data protection and transparency while ensuring accountability among developers and institutional actors [118, 122]. Future studies should clarify how such ethical principles can be operationalized in high-stakes fields like healthcare and academic publishing, ensuring that GenAI tools uphold the expected integrity and respect data sovereignty. The proposal of a modern Turing test to detect AI-generated research submissions further highlights the urgency of maintaining integrity in scholarly communication [126]. As GenAI is increasingly used in decision systems, including the area of security, investigators should also strive to address biases in GenAI-powered fraud detection systems which can result in higher false positive rates for transactions from certain demographic groups, leading to discriminatory practices [38].

**Demonstrating trustworthiness.** Transparency is considered fundamental for demonstrating GenAI's trustworthiness, especially in healthcare where the need for interpretable models that clearly communicate decision-making processes is crucial to foster clinician acceptance [8, 111, 114]. Explainability mechanisms are equally necessary for building confidence among users, especially healthcare professionals, particularly as design science researchers develop new methods to enhance these capabilities and study their effects on adoption [8, 121]. Reliability and accuracy improvements cannot be ignored either. Studies should evaluate GenAI's performance across broader domains, e.g., medical specialties which require complex diagnostic and treatment decisions where accuracy is extremely important [114, 115]. This could be supported with simplifying model architectures, implementing standardized validation procedures, and addressing data quality challenges to ensure models meet clinical standards [119, 133]. Mitigating hallucinations and inaccuracies in LLM outputs demands urgent attention as these issues can undermine trust [8, 124]. Researchers should strive to develop techniques beyond Chain-of-Thought reasoning to mitigate inaccuracies and improve verifiability [124, 130].

Human-AI interaction research should investigate user attitudes toward GenAI, examining why people appreciate or avoid these tools and exploring trust-related factors [8, 112]. This includes studying impacts on business workers and general users, requiring renewed focus on end-user computing topics in the GenAI era [35]. Design approaches that foster trust through improved system reliability and transparency will be essential [121]. Future work should also address bias issues affecting GenAI integration, e.g., in clinical workflows [112]. Understanding the factors that influence user trust in LLM outputs – despite potential hallucinations and training data inaccuracies – is essential [8].

**Efficiency and scalability.** Future research on GenAI efficiency and scalability encompasses several interconnected dimensions. Among them, operational integration emerges as a critical theme, focusing on how GenAI can transform digital services through enhanced efficiency and operational capabilities [130]. This includes investigating government adoption strategies for building scalable digital services accessible to broader populations [130]. Parallel to service delivery, the localization domain demonstrates significant potential, where GenAI integration enhances efficiency, scalability, and cultural customization, enabling hyper-localized content creation [125]. Quality assurance and reliability represent another essential research direction, particularly in high-stakes applications. Prioritizing model accuracy, reliability, and robustness ensures safe and effective clinical applications [114]. This intersects well with bias mitigation challenges, prompting researchers to design dynamic methodologies that integrate real-time user feedback for continuous detection and mitigation of LLM bias while maintaining scalability and operational efficiency [38]. Beyond individual organizational contexts, inter-organizational dynamics call for systematic investigation. As GenAI permeates organizations, interaction patterns between and among entities will evolve, necessitating research into efficient cooperation across organizational boundaries [35].

**Accountability and contestability.** A valid research challenge lies in balancing competing priorities: high-speed decision-making versus ethical accountability, and hyper-personalization versus inclusivity [38]. In this context, operationalizing bias mitigation is important yet requires transparent, explainable interfaces that preserve human agency while implementing robust safeguards [117]. Research should clarify the obligations of GenAI providers to determine which GenAI applications should face restrictions or prohibition [128]. This involves establishing clear accountability frameworks that delineate responsibilities among developers, providers, and regulators [117]. Another identified research direction is addressing the challenge of responsibly automating organizations' business processes while ensuring transparency and accountability in GenAI deployment [130]. Future work should also advance reliability through information resilience protocols, including security measures against prompt injection and data poisoning [117]. Essential components include human-in-the-loop oversight and contestability mechanisms that enable stakeholders to challenge AI-driven outcomes [117].

## 7. Discussion

GenAI is transitioning from a standalone technological novelty to a core constituent of modern socio-technical systems. Our findings indicate that its transformative potential is not an inherent property of the models themselves, but rather an emergent outcome of the deliberate alignment between artifacts, human expertise, organizational processes, and institutional frameworks. We argue that by reframing GenAI as a socio-technical entity, we can better understand the variance in model performance across different domains and recognize that governance, workflow design, and human oversight are as critical as algorithmic precision. This perspective shifts the focus from simple tool adoption toward the design of hybrid human-AI ensembles, i.e., systems where value is contingent upon the collaborative interaction between human intelligence and machine capability.

The synthesis of findings across diverse sectors reveals that GenAI value materializes primarily through complementarity rather than the wholesale substitution of human labor. For example, within a healthcare context, ambient AI documentation systems have demonstrated significant reductions in administrative burden and after-hours work (provided that the workflow maintains rigorous human review and safety-case attestation). Similarly, in software engineering, AI-augmented programming accelerates routine tasks and facilitates knowledge diffusion particularly among less experienced developers. These gains are maximized when organizations implement clear role allocations, such as the assistant–reviewer–attestor triad, and maintain robust provenance of AI-generated contributions. These findings align with broader productivity studies suggesting that the benefits of GenAI accrue disproportionately to those who leverage it for hybrid learning and skill augmentation. Yet this very complementarity creates a distinctive socio-technical contradiction, i.e., the more successfully GenAI lowers cognitive load and entry barriers, the greater the risk of human automation bias and long-term erosion of critical judgment (the exact capabilities required to supervise the system itself).

Despite these benefits, our findings indicate that the transition from successful laboratory pilots to sustained field impact remains fraught with socio-technical challenges. Inconsistencies in outcomes often stem from a lack of "fit" rather than technical failure. For example, while GenAI enables hyper-personalization in education and marketing, its success depends on pre-existing AI literacy, redesigned assessment frameworks and stringent accessibility standards. Furthermore, the persistent challenges of representational harm and cultural bias highlight how models encode latent societal tendencies, requiring participatory governance to prevent marginalization. Reliability also remains a significant hurdle. Fixed training cut-offs produce widening knowledge gaps. Hallucinations also yield outputs that are plausible yet wrong. Our findings indicate that limited contextual awareness yields formally correct but practically dangerous outputs, and performance drift silently alters behavior after deployment. Together, these failure modes strike at the IS discipline's foundational commitments to system dependability, traceability, and accountability and transforms apparent technical limitations into a full-blown crisis of socio-technical legitimacy.

Addressing these risks requires a shift toward defense-in-depth security architectures and risk-based governance frameworks, such as the NIST AI Risk Management Framework [135], to manage the emerging threats of prompt injection and shadow AI while preserving the human subsystem as the ultimate locus of responsibility and control.

### 7.1. The sociotechnical GenAI outcomes matrix (SGOM)

This section introduces the Sociotechnical GenAI Outcomes Matrix (SGOM). SGOM (Table 12) provides a conceptual framework for synthesizing the evidence gathered in this study. It posits that GenAI outcomes are co-produced across multiple levels of analysis, ranging from individual users to technical engineering layers.

To interpret this matrix, each row provides a distinct socio-technical dimensions that must be synchronized for successful implementation. The columns represent a progression from empirical observation ("What works"/"Why it fails") to normative intervention ("Design moves") and finally to evaluative rigor ("Evidence/KPIs"). Effective GenAI deployment requires moving horizontally across a row to ensure that every benefit is protected by a corresponding control and measured by a domain-specific metric. Conversely, vertical alignment ensures that technical engineering controls (e.g., RAG hardening) support higher-level ethical and organizational goals.

The SGOM serves as a diagnostic instrument that nudges IS research away from monolithic evaluations of model "intelligence" toward a nuanced understanding of situated performance. By mapping technical failures (e.g., performance drift) directly to organizational controls (e.g., accountability maps), the matrix forces researchers to move beyond the black box view of GenAI. It also provides a structured vocabulary to describe the interdependencies between subsystems and illustrates, for example, how societal biases are not merely data errors but governance failures that require participatory design. Consequently, we contend that the SGOM acts as a roadmap for future research, encouraging scholars to investigate not just whether a model works, but under what specific socio-technical configurations its benefits become durable and legitimate.

Table 12. Sociotechnical GenAI outcomes matrix (SGOM)

| Perspective | What works (benefits) | Why it fails (challenges) | High-value design moves (controls) | Evidence/KPIs |
|---|---|---|---|---|
| User (e.g., clinicians, educators, developers) | Hybrid human–AI teaming improves throughput and quality via role clarity and calibration. It also reduces professional burnout and cognitive load. | Over-reliance; contextual blind spots; knowledge staleness; literacy gaps. | Human-in-the-loop attestation; literacy programs; Retrieval-Augmented Generation (RAG) freshness with citations | Task time reduction; correction rates; user satisfaction (SPACE). |
| Organizational (e.g., workflows, processes) | Standardized artifact provenance and ensemble design enable knowledge diffusion. | Lack of evaluation frameworks; pilot-to-production attrition; brittle governance. | Assistant – reviewer – attestor patterns; evidence ladders; accountability maps. | Time to insight; defect density; conversion from proof of concept to production. |
| Societal (e.g., culture, equity, public services) | Democratizes access to specialized knowledge. Multilingual access and personalization at scale improve service participation. | Representational harms; latent cultural tendencies; accessibility gaps; environmental costs of model training and operation. | Participatory governance; cross-cultural evaluation; content watermarking; carbon-aware deployment policies. | Inclusion indices; bias audits; provenance coverage; carbon footprint assessment. |
| Ethical (e.g., fairness, transparency) | Reason-giving transparency (explaining "why") sustains trust and enables redress. | Opaque behaviors; immature XAI; lack of clear liability for harms; unresolved IP ownership disputes. | Model/data cards; explainability UIs; contestability and incident registers; IP clearance protocols. | Audit pass rates; explainability adequacy; time to redress; IP dispute rates; attribution coverage. |
| Engineering (e.g., security, robustness) | Defense-in-depth and validated pipelines (repair/validation loops) improve safety. | Prompt injection; performance drift; limited contextual awareness. | OWASP LLM Top 10; red teaming; RAG hardening; canary tests; drift dashboards. | Attack block rate; jailbreak detection; semantic robustness score. |
| Quality (e.g., evaluation, safety) | Efficiency gains materialized via validated frameworks (e.g., PDQI-9). | Evidence gaps; weak metrics; reproducibility issues; lack of public datasets. | Standardized domain metrics; multi-site benchmarking; error taxonomies. | PDQI-9 scores; hallucination rate; inter-rater agreement; benchmark pass rates. |

### 7.2. Implications for research and practice

From a Socio-Technical Systems (STS) theory perspective, the introduction of GenAI into IS represents a fundamental shift in the joint optimization of the social and technical subsystems. Our research implies that the primary unit of analysis in IS must evolve from the individual tool or the isolated user to the hybrid human-AI ensemble.

#### 7.2.1. Implications for research

Researchers should move beyond simple adoption models to explore how GenAI alters organizational structures and power dynamics. This necessitates the development of formal constructs for ensemble coordination, specifically focusing on how role allocation (assistant vs. attestor) influences the variance in work outcomes. Applying the lens of Structural Contradiction, future work should investigate the tension between the technical efficiency of GenAI and the social requirement for accountability. Furthermore, the IS field must lead in establishing 'evidence ladders', i.e., methodological frameworks that transition from simulated model testing to high-fidelity field trials. This will ensure that "social fit" is measured with the same rigor as technical accuracy.

#### 7.2.2. Implications for practice

For practitioners, the socio-technical perspective mandates that GenAI deployment is treated as an organizational redesign rather than a software upgrade. Managers must prioritize the "secondary design" of the social subsystem. This includes investing in AI literacy and new job descriptions to match the capabilities of the technical subsystem. This also involves operationalizing risk-based governance (e.g., NIST AI RMF [135]) not as a compliance checklist, but as a dynamic mechanism for maintaining institutional legitimacy. Finally, practitioners must implement "contestability by design", providing human users with the technical tools to adjudicate, override, and audit AI outputs. This will help to ensure that the human subsystem remains the ultimate locus of responsibility in high-stakes environments.

Our research indicates that the frontier of GenAI is fundamentally socio-technical. The ultimate value of these systems is not derived from the raw power of the underlying models, but from the sophistication of the ensembles in which they are embedded. By aligning technical advances with institutional legitimacy, standardized field metrics, and risk-based governance, GenAI can transition from an experimental technology to a durable, trustworthy pillar of modern IS.

## 8. Future research agenda

Our study highlights transformative benefits across domains, but also a triad of constraints – technical unreliability, pervasive societal and ethical risks, and a systemic governance vacuum – that together signal a persistent misalignment between GenAI's fast-evolving technical subsystem and the slower-adapting social and institutional arrangements in which it is embedded. GenAI capabilities are advancing faster than the norms, governance structures, and regulatory institutions required for responsible deployment. Consequently, the path forward is not merely a holistic investigation but demands proactive intervention and socio-technical design aimed at achieving joint optimization. The SGOM operationalizes this design challenge, providing a scaffold that links observed failure modes to the necessary technical, social, organizational, and institutional controls – and the KPIs required to evaluate whether performance becomes durable in practice. Building on this logic, we articulate a research agenda that reorients IS scholarship from analyzing impacts toward actively shaping the co-evolution of these interdependent subsystems, organized around three critical frontiers.

### 8.1. Frontier 1: Organizational reconfiguration and governance

The "governance vacuum" identified in our results indicates that traditional IT governance structures are ill-equipped for the decentralized, pervasive nature of GenAI.

**Governing "shadow AI".** Unlike traditional enterprise software, GenAI is easily accessible to individual employees, leading to unsanctioned use. Research is needed to develop governance frameworks that balance the innovation potential of bottom-up adoption with the risks of data leakage, privacy violations, and regulatory non-compliance.

**New workflows and role definitions.** The transition to hybrid human-AI collaboration requires revisiting organizational routines. Research should explore how accountability is distributed in the "assistant–reviewer–attestor" triad. Who is liable when an AI-generated, human-reviewed artifact fails? How must job descriptions evolve to prioritize "verification skills" over "creation skills"? How do organizations redesign business processes to integrate GenAI while maintaining transparency and human oversight. More broadly, this transition points toward the emergence of new occupational categories and labor market structures shaped by human–AI collaboration.

**Regulatory translation.** How do emerging regulatory frameworks (e.g., the EU AI Act, NIST AI RMF) translate into organizational governance practices? Comparative studies are needed to understand the barriers to effective implementation of these external mandates within internal workflows.

### 8.2. Frontier 2: Societal alignment, ethics, and law

Our findings on bias and representational harm confirm that GenAI is not culturally neutral. The agenda here should move from problem identification to solution engineering informed by socio-technical perspectives.

**Human-AI symbiosis.** Future research should empirically investigate the boundary between helpful augmentation and harmful dependency. Longitudinal studies are needed to measure how relying on GenAI for coding, writing, or diagnostics affects human domain expertise, critical thinking, and professional identity over time. How do we design interactions that maintain "human-in-the-loop" vigilance without causing fatigue or the erosion of tacit knowledge? IS scholarship can extend this by theorizing "trust calibration" and "appropriate reliance" in routine work, knowledge work, and high-stakes decision support.

**Operationalizing fairness and algorithmic justice.** While the literature identifies bias as a major risk, there is a scarcity of frameworks for operationalizing fairness in specific industries. Research should focus on developing domain-specific audit protocols (e.g., for healthcare triage) that align algorithmic outputs with local legal and ethical standards.

**Participatory design.** To counter "exclusionary norms", IS researchers should lead participatory design initiatives that involve marginalized communities in the fine-tuning and evaluation of models, ensuring that GenAI systems reflect diverse cultural and linguistic realities rather than just dominant training data.

**Intellectual property and value attribution.** As GenAI disrupts the economics of knowledge production, research is needed into new legal and economic models for attributing value. How can we trace provenance in AI-generated content to ensure fair compensation for original creators? IS research should examine the implications for digital platforms and content ecosystems, investigating how provenance-tracking technologies can enable transparent attribution. Studies should analyze the economic sustainability of creative industries in the GenAI era, the emergence of new intermediaries and market mechanisms for AI-generated content. Research drawing on platform governance and digital rights management literature can inform the design of attribution systems that balance creator rights with innovation incentives.

**Information integrity and authenticity in digital ecosystems.** Research should explore socio-technical safeguards against AI-generated misinformation (verification routines, provenance signals that users can interpret, moderation policies, and institutional responses).

### 8.3. Frontier 3: Design and validation of GenAI artifacts

The probabilistic and generative nature of GenAI challenges the deterministic assumptions often held in IS design and evaluation. To address the "reliability crisis" identified in our findings, IS scholars must spearhead a research program focused on the rigorous design and contextual validation of these artifacts within organizational settings.

**Design principles for probabilistic systems.** Traditional IS design theory often assumes consistent system behavior. Future IS research must formulate new design principles and meta-requirements for systems that are inherently unstable or prone to hallucination. How do we design IT artifacts that remain useful and trustworthy even when the underlying model is imperfect? This includes design for uncertainty (calibration cues, confidence communication, and verification affordances), and explicit "safety cases" that connect model limitations to workflow controls.

**Designing for contestability.** Reliability requires that users can challenge AI outputs. Design Science Research should focus on creating interfaces that support "contestability by design" – mechanisms that allow users to easily audit and query model outputs, shifting the user role from passive consumer to active auditor.

**Secure-by-design and privacy-by-design GenAI.** Building on prompt-injection/jailbreak and privacy-leakage concerns, future work should design and evaluate defense-in-depth patterns for GenAI-enabled IS, and examine how these technical controls interact with organizational routines and user practices.

**Green IS and corporate digital responsibility.** Aligning with the discipline's growing focus on sustainability, researchers should investigate the trade-offs between model performance and environmental impact. IS scholars are positioned to develop decision frameworks for "Green GenAI", helping organizations balance the computational costs (energy and financial) of LLMs against their actual business value, and promoting the adoption of "frugal AI" strategies where appropriate.

**Socio-technical evaluation frameworks.** Traditional metrics are insufficient for evaluating GenAI. Research should develop domain-specific evaluation frameworks that capture helpfulness, harmlessness, and contextual appropriateness.

By pursuing this agenda, the IS community can fulfill its critical role as the bridge between the technical frontier of AI development and the social, organizational, and ethical contexts in which these systems must operate. The challenge before us is to leverage this distinctive positioning to ensure that GenAI evolves as a technology that augments human capability, respects human dignity, and serves the broad public interest.

## 9. Threats to validity

As with any systematic review, this study is subject to limitations that must be considered when interpreting the findings. We discuss these threats to validity following the classification framework for secondary studies proposed by [102].

### 9.1. Study selection validity

This category concerns the risk of missing relevant studies or selecting inappropriate ones.

- **Selection of digital libraries.** While we queried three premier repositories for IS research (AIS eLibrary, Scopus, and Web of Science), we acknowledge that extending the search to additional databases (e.g., Springer, Wiley, Emerald, and Taylor and Francis) could have improved coverage. However, Scopus and Web of Science index papers from multiple publishers, which partially mitigates this limitation.
- **Search strategy limitations.** The search string was systematically constructed using three facets (study type, phenomenon, and domain) and incrementally revised by multiple co-authors. However, some studies using non-standard terminology may have been missed.
- **Selection of arbitrary starting year.** We restricted our search to publications from 2023 onwards to capture the post-ChatGPT surge in GenAI research. This improves topical focus but may under-represent earlier IS-relevant work on generative models and may overweight LLM-centric framings that became dominant after late 2022.
- **Subjectivity in screening.** Despite employing a multi-stage screening process with independent reviewers and consensus meetings, the interpretation of inclusion and exclusion criteria inherently involves subjective judgment. Although all conflicts were resolved through consensus, different research teams might arrive at marginally different sets of included studies.
- **Exclusion of grey literature.** Consistent with our focus on peer-reviewed secondary studies, we excluded grey literature. As for preprints, significant insights may be disseminated through non-traditional channels before appearing in peer-reviewed venues. As for technical reports, white papers, and industry publications, we believe that they are not appropriate to deliver literature review studies. However, this choice may exclude influential practitioner roadmaps that shape GenAI governance and adoption in IS practice, potentially under-representing practice-led developments.
- **Temporal currency of search.** The systematic search was conducted in April 2025, thus excluding literature published thereafter from our analysis. Given the exceptionally rapid pace of GenAI development and the continuous emergence of new research, our findings may not reflect the most recent advances in technical capabilities, governance frameworks, or application domains. However, this temporal lag is an inherent and well-recognized trade-off in rigorous systematic literature reviews, as comprehensive screening, quality assessment, and thematic synthesis demand substantial time. Despite this limitation, our core socio-technical findings and proposed research agenda remain robust and highly relevant to the IS community.

### 9.2. Data validity

This category concerns the validity of the extracted dataset and its analysis.

- **Data extraction reliability.** Data extraction was conducted by six reviewers following a structured protocol refined during a pilot phase. While weekly alignment meetings and quality audits were employed to ensure consistency, the extraction of qualitative dataparticularly the identification of benefits, challenges, and future research directions – required interpretive judgment. One reviewer's work was identified as deficient during quality audits and was subsequently re-evaluated by another team member. Although this mitigation strategy improved data quality, it highlights the inherent challenges of maintaining consistency across multiple extractors. To support auditability, we captured verbatim quotations during extraction, enabling traceability from synthesized themes back to the source texts.
- **Validity of secondary studies.** As a review of secondary studies, our findings are contingent upon the rigor and accuracy of the included reviews and research agendas. Any errors, biases, or omissions present in these source documents are transitively reflected in our synthesis. We did not verify the primary studies underlying the included secondary studies. Consequently, the strength of our conclusions is bounded by the quality of the evidence base we inherited. We mitigated this by excluding non-peer-reviewed sources.
- **Quality assessment limitations.** While we applied formal quality assessment criteria (DARE for secondary studies, custom criteria for research agenda papers), quality assessment inherently involves subjective judgment. Different reviewers might assign marginally different ratings to the same study. We mitigated this through dual-reviewer assessment with consensus resolution and reported inter-rater agreement statistics. Additionally, we did not exclude studies based on quality ratings. This decision was made to ensure comprehensive coverage of the nascent GenAI literature, where even methodologically imperfect studies may offer valuable insights. However, this approach means that the quality of evidence synthesized in our review varies across included studies. To support informed interpretation, we report individual study quality scores alongside each synthesized finding in Tables 9, 10, and 11, enabling readers to gauge the evidential strength of specific findings and to interpret them with this heterogeneity in mind.
- **Heterogeneity of source material.** Our review includes *secondary studies* and *research agendas*. While this provides a holistic view of the field, it introduces heterogeneity in the granularity of evidence. SLRs typically provide retrospective empirical evidence, while agendas provide prospective theoretical propositions. To address this, we used a flexible thematic synthesis method capable of handling diverse qualitative inputs.
- **Risk of double counting.** When synthesizing findings across multiple secondary studies, there is a risk that overlapping primary studies may be counted multiple times, potentially inflating the apparent strength of certain findings. While we did not conduct an overlap analysis of primary studies across included reviews, researchers should be aware of this limitation when interpreting the prevalence of specific themes.

### 9.3. Research validity

This category concerns the analysis procedures and the reproducibility of the study.

- **Interpretive bias in thematic analysis and synthesis.** The qualitative analysis and synthesis of benefits, challenges, and future directions utilized a Grounded Theory-inspired approach, which is inherently interpretive. To support transparency, we documented our research protocol in detail and made all intermediate artifacts – including screening decisions, extraction sheets, and coding outputs – publicly available in the online replication package. However different researchers applying the same protocol may arrive at somewhat different thematic structures, category labels, or emphasis in synthesis. To mitigate this, we employed a multi-analyst approach with independent coding and consensus meetings.
- **Generalizability.** Our findings are bounded by the Information Systems discipline and the tertiary nature of our review. As we synthesize secondary studies, our results are constrained by what the included reviews chose to report, the application sectors and contexts they covered, and the nascent state of a rapidly evolving field. Consequently, these insights may not generalize to domains or industries not explicitly covered by the included secondary studies.

## 10. Conclusions

This study synthesizes a rapidly evolving body of Information Systems research on Generative AI (GenAI), drawing on qualified sources published predominantly in 2024 and 2025. The literature, mainly published in journal outlets, spans core IS application areas, with the strongest coverage in healthcare and social work contexts, followed by information and communication, professional and technical services, and education. Our analysis grouped the findings into three themes: benefits, challenges and limitations, and research gaps and future directions.

Regarding benefits, the reviewed literature highlights improvements to information work in practice. Specific advancements include enhanced diagnostic accuracy, accelerated drug discovery, and streamlined clinical documentation in healthcare; more personalized and accessible learning with reduced educator workloads in education; augmented knowledge synthesis and creative prototyping in research and design; productivity gains in software engineering through automated coding, testing, and translation support; and the delivery of scalable, personalized digital services. These advantages also extend to more efficient data management, multimodal content generation, and synthetic data creation, which are positioned to support privacy-preserving and ethical AI initiatives.

Nonetheless, the same body of evidence demonstrates that GenAI deployment is constrained by a complex landscape of challenges and limitations spanning four primary dimensions. First, societal and ethical challenges include the amplification of biases, the spread of misinformation, human skill degradation, and high environmental costs. Second, technical unreliability is driven by hallucinations, limited contextual awareness, performance drift, and a lack of empirical validation. Third, the opacity and explainability shortcomings of "black-box" architectures obstruct system auditing and user trust. Finally, a persistent governance and legal vacuum exposes organizations to data privacy breaches, security threats, intellectual property disputes, and accountability ambiguities.

To integrate these insights, we introduced the Sociotechnical GenAI Outcomes Matrix (SGOM) as a conceptual framework that links observed benefits and challenges to socio-technical dimensions. The SGOM reframes GenAI outcomes as co-produced across user, organizational, societal, ethical, engineering, and quality perspectives. This framing has two implications that matter for IS scholarship and practice. For research, it motivates a shift from studying isolated tool adoption or model performance toward studying the hybrid human-AI ensemble. For practice, it emphasizes that the implementation of GenAI constitutes an organizational redesign challenge, necessitating explicit controls, traceability, and contestability mechanisms to uphold human responsibility.

Future research in Information Systems should prioritize understanding the socio-technical conditions that ensure the safe and accountable use of GenAI, rather than simply documenting its immediate effects. This includes investigating how organizations govern decentralized "shadow" AI use, how ethical and legal requirements are translated into everyday routines, and how IT artifacts developed with GenAI can be both useful and trustworthy, while also allowing for human contestability.

## Acknowledgment

This paper is dedicated to the memory of Professor Stanisław Wrycza, the visionary founder of the International Conference on Information Systems Development (ISD). The collaboration underlying this study was initiated during the 32nd edition of the ISD conference in 2024, a testament to his enduring legacy in fostering a vibrant and innovative IS research community.

## CRediT authorship contribution statement

A. Jarzębowicz: conceptualization, methodology, investigation, data curation, writing – original draft, writing – review and editing, project administration. A. Przybyłek: conceptualization, methodology, software, investigation, data curation, writing – original draft, writing – review and editing, project administration. J. Estima: methodology, investigation, data curation, writing – original draft. Y. Y. Ng: investigation, data curation, writing – original draft. J. Swacha: investigation, data curation, writing – original draft. B. Zielosko: investigation, data curation, writing – original draft, visualization. L. Madeyski: methodology, software, writing – original draft, writing – review and editing. N. Carroll: writing – original draft. K.-K. Kemmell: writing – original draft. B. Marcinkowski: writing – original draft, visualization. A. Rodrigues da Silva: writing – original draft, writing – review and editing. V. Stray: writing – original draft, writing – review and editing. N. Iivari: writing – original draft. A. Nguyen-Duc: writing – original draft. J. Melegati: writing – original draft. B. Delibašić: writing – original draft. E. Insfran: writing – review and editing.
The detailed description of authors' contributions is available at:
https://github.com/przybylek/GenAI4IS

## Declaration of competing interest

The authors declare that they have no competing interests.

## Data and code availability

The replication package is available at https://github.com/przybylek/GenAI4IS. It includes the detailed thematic codebooks – containing verbatim excerpts from each included study mapped to their respective codes – as well as screening decisions, pre-consensus reviewer ratings with justifying comments, data extraction sheets, and analysis scripts.

## Funding

This research was partially supported by the University of Belgrade – Faculty of Organizational Sciences and, in part, by the Ministry of Science, Technological Development and Innovation of the Republic of Serbia through institutional funding (grant number: 200151).
This work was supported, in part, by Taighde Éireann – Research Ireland under Grant number 13/RC/2094_2. Co-funded by the European Union under the Systems, Methods, Context (SyMeCo) Programme Grant Agreement Number 101081459. Views and opinions expressed are however those of the author(s) only and do not necessarily reflect those of the European Union or the European Research Executive Agency. Neither the European Union nor the granting authority can be held responsible for them.
This work was partially supported by the Europium Short-Term Outgoing Visits program from Gdańsk University of Technology's Initiative of Excellence.
This work was partially funded by national funds through FCT – Foundation for Science and Technology, I.P., within the scope of the research unit UID/00326 – Centre for Informatics and Systems of the University of Coimbra, https://doi.org/10.54499/UID/00326/2025.

## Declaration of AI assistance in the writing process

The authors employed Large Language Models (LLMs) to support the thematic analysis, specifically to brainstorm initial code names and propose refined labels and concise descriptions for the inductively developed codes and themes. In all cases, LLM suggestions served only as initial inspiration; many were discarded for being too narrow or conflating distinct concepts. All final code names, theme structures, and descriptions were critically evaluated and approved by the research team. Additionally, LLMs were used to improve the language, clarity, and readability of selected passages. All AI-generated suggestions were reviewed and edited by the authors, who take full responsibility for the final content of this paper.

## Authors and affiliations

Aleksander Jarzębowicz
e-mail: aleksander.jarzebowicz@pg.edu.pl
ORCID: https://orcid.org/0000-0003-3181-4210
Department of Software Engineering, Gdańsk University of Technology, Poland

Adam Przybyłek
e-mail: adam.przybylek@gmail.com
ORCID: https://orcid.org/0000-0002-8231-709X
Department of Software Engineering, Gdańsk University of Technology, Poland; J.E. Cairnes School of Business and Economics, University of Galway, Ireland; Lero, the Research Ireland Centre for Software, Ireland

Jacinto Estima
e-mail: estima@dei.uc.pt
ORCID: https://orcid.org/0000-0001-8837-4637
Department of Informatics Engineering, CISUC/LASI, University of Coimbra, Portugal

Yen Ying Ng
e-mail: nyysang@hotmail.com
ORCID: https://orcid.org/0000-0001-5388-2025
Center for Language Evolution Studies, Nicolaus Copernicus University in Toruń, Poland

Jakub Swacha
e-mail: jakub.swacha@usz.edu.pl
ORCID: https://orcid.org/0000-0002-2214-6989
Department of IT in Management, University of Szczecin, Poland

Beata Zielosko
e-mail: beata.zielosko@us.edu.pl
ORCID: https://orcid.org/0000-0003-3788-1094
Institute of Computer Science, University of Silesia in Katowice, Poland

Lech Madeyski
e-mail: Lech.Madeyski@pwr.edu.pl
ORCID: https://orcid.org/0000-0003-3907-3357
Faculty of Information and Communication Technology, Wroclaw University of Science and Technology, Poland

Noel Carroll
e-mail: noel.carroll@universityofgalway.ie
ORCID: https://orcid.org/0000-0001-8344-1220
J.E. Cairnes School of Business and Economics, University of Galway, Ireland

Kai-Kristian Kemell
e-mail: kai-kristian.kemell@tuni.fi
ORCID: https://orcid.org/0000-0002-0225-4560
Faculty of Information Technology and Communication Sciences, Tampere University, Finland

Bartosz Marcinkowski
e-mail: bartosz.marcinkowski@ug.edu.pl
ORCID: https://orcid.org/0000-0002-7230-2978
Department of Business Informatics, University of Gdańsk, Poland

Alberto Rodrigues da Silva
e-mail: alberto.silva@tecnico.ulisboa.pt
ORCID: https://orcid.org/0000-0002-7900-9846
INESC-ID, Instituto Superior Técnico, University of Lisbon, Portugal

Viktoria Stray
e-mail: stray@ifi.uio.no
ORCID: https://orcid.org/0000-0002-6032-2074
Department of Informatics, University of Oslo, Norway

Netta Iivari
e-mail: Netta.Iivari@oulu.fi
ORCID: https://orcid.org/0000-0002-7420-2890
INTERACT Research Group, University of Oulu, Finland

Anh Nguyen-Duc
e-mail: anh.nguyen.duc@usn.no
ORCID: https://orcid.org/0000-0002-7063-9200
Department of Business and IT, University of South Eastern Norway, Norway

Jorge Melegati
e-mail: melegati@fe.up.pt
ORCID: https://orcid.org/0000-0003-1303-4173
INESC TEC, Faculty of Engineering, University of Porto, Portugal

Boris Delibašić
e-mail: boris.delibasic@fon.bg.ac.rs
ORCID: https://orcid.org/0000-0002-6153-5119
Faculty of Organizational Sciences, University of Belgrade, Serbia

Emilio Insfran
e-mail: einsfran@dsic.upv.es
ORCID: https://orcid.org/0000-0003-0855-5564
Department of Software Systems and Computation, Universitat Politècnica de València, Spain